\makeatletter
\declare@file@substitution{revtex4-1.cls}{revtex4-2.cls}
\makeatother

\documentclass[twocolumn,trackchanges,twocolappendix]{aastex631}

\usepackage[colorinlistoftodos]{todonotes}               
\usepackage{amsmath}
\usepackage{colortbl}
\usepackage{multirow}
\usepackage{natbib}

\newcommand{\cm}{\mathrm{cm}}
\newcommand{\g}{\mathrm{g}}
\newcommand{\G}{\mathrm{G}}
\newcommand{\K}{\mathrm{K}}  
\newcommand{\km}{\mathrm{km}}

\newcommand{\pc}{\mathrm{pc}}

\newcommand{\erg}{\mathrm{erg}}

\newcommand{\s}{\mathrm{s}}

\newcommand{\incgraph}[1]
  {\includegraphics[width=\linewidth]{#1}}
\newcommand{\kratos}{KRATOS }

\begin{document}

\title{Cloud Crushing and Dissipation of Uniformly-Driven Adiabatic Turbulence in Circumgalactic Media}

\author[0009-0002-8164-7975]{Alex Lv}
\affiliation{Kavli Institute for Astronomy and Astrophysics,
  Peking University, Beijing 100871, People's Republic of
  China}
\affiliation{Department of Astronomy, School of Physics,
  Peking University, Beijing 100871, People's Republic of
  China}
\email{alexanderlu@stu.pku.edu.cn}

\author[0000-0002-6540-7042]{Lile Wang}
\affiliation{Kavli Institute for Astronomy and Astrophysics,
  Peking University, Beijing 100871, People's Republic of
  China}
\affiliation{Department of Astronomy, School of Physics,
  Peking University, Beijing 100871, People's Republic of
  China}
\correspondingauthor{Lile Wang} \email{lilew@pku.edu.cn}

\author[0000-0001-8531-9536]{Renyue Cen}
\affiliation{Institute for Advanced Study in Physics,
  Zhejiang University, Hangzhou 310027, People's Republic of
  China}
\affiliation{Institute of Astronomy, School of Physics, Zhejiang University, Hangzhou 310027, People’s Republic of China}
\email{renyuecen@zju.edu.cn}

\author[0000-0001-6947-5846]{Luis C. Ho}
\affiliation{Kavli Institute for Astronomy and Astrophysics,
  Peking University, Beijing 100871, People's Republic of
  China}
\affiliation{Department of Astronomy, School of Physics,
  Peking University, Beijing 100871, People's Republic of
  China}
\email{lho.pku@gmail.com}

\begin{abstract}
    The circumgalactic medium (CGM) is responsive to
    kinetic disruptions generated by
    nearby astrophysical events. In this work, we study the
    saturation and dissipation of turbulent hydrodynamics
    within the CGM through an extensive array of 252
    numerical simulations with a large parameter space. These simulations are
    endowed with proper cooling mechanisms to consistently
    explore the parameter space spanned by the average gas
    density, metallicity, and turbulence
    driving strength. A dichotomy emerges in the
    dynamics dissipation behaviors. Disturbances that are
    hot and subsonic are characterized by weak compression
    and slow dissipation, resulting in density
    fluctuations typically $\lesssim 10^{-2}$. Conversely,
    warm supersonic turbulence, marked by significant
    compression shocks and subsequent rapid cooling, is associated
    with substantial clumping factors $\sim 10^0-10^1$.  In
    the supersonic cases, the kinetic energy decay is
    divided into a rate-limiting phase of shock dissipation
    and a comparatively swift phase of thermal dissipation,
    predominantly occurring within the overdense regions. Upon turbulence driving turnoff, the
    strong density contrasts decay within a
    relatively brief timescale of $\sim 30 - 300~{\rm Myr}$,
    depending on the average gas density. Dense clouds are crushed on similar 
    timescales of $ \sim 30 - 100 ~{\rm Myr}$, depending on turbulence driving
    strength but independent from 
    average gas density. Results of this
    work also contribute a novel dataset of dissipation
    timescales that incorporates an understanding of kinematics and thermodynamics in
    addition to the traditional cooling rate tables, which
    may serve as a valuable asset for forthcoming
    simulations that aim to explore gas dynamics on
    galactic and cosmological scales.
\end{abstract}

\keywords{Circumgalactic medium (1879) --- Collapsing clouds (267) --- Hydrodynamical simulations (767) --- Extragalactic astronomy (506)}

\section{Introduction} \label{sec:intro}

The circumgalactic medium (CGM) represents a rich,
multiphased environment surrounding a galaxy, extending out
up to the virial radius of the halo of $10^1 - 10^2$ kpc. It
plays a crucial role in galaxy evolution, serving as an
important source of baryons \citep{2014ApJ...792....8W} for
gas accretion onto the galactic disc. The physical state of
the CGM can be split into distinct phases
\citep{2013ApJ...770..139C}: The cool phase of the CGM
typically hovers around $10^4$ to $10^5$ K, while the hot
phase $10^5$ to $10^6$ K \citep{2012ApJ...750...67R,
  2013ApJS..204...17W}, meaning the bulk of the CGM consists
of ionized gas. The mode of accretion depends heavily on the
cooling timescale of the gas in relation to the freefall
time: cold mode accretion characterized by filamentary
streams and clumps when $t_\mathrm{cool} < t_\mathrm{ff}$
and hot mode accretion characterized by smooth cooling flows
when $t_\mathrm{cool} > t_\mathrm{ff}$
\citep{1978MNRAS.183..341W,2013ApJ...765...89S,
  2013ApJ...776L..18C, 2020MNRAS.492.6042S}). However, given
the multiphased nature of the CGM, with varying
metallicities, ionization states and temperatures,
calculating and characterizing cooling timecales is a highly
complex and nonlinear problem. Generally speaking, cooling
streams only occur in the non-virialized regions of the
circumgalactic medium where the accretion rate exceeds a
critical threshold \citep{2023ARA&A..61..131F}. Thermal
instabilities due to rapid cooling shocks the gas, leading
to the formation of cold supersonic filaments
\citep{1989ApJ...341..611B,2009ApJ...703..785D,2021ApJ...911...88S}.

Unlike typical giant molecular cloud (GMC) conditions in the
interstellar medium (ISM), in addition to being much hotter
and mostly ionized, the CGM is also more diffuse.
Observations reveal typical hydrogen column densities
ranging from $10^{20} \; \mathrm{cm}^{-2}$ near the disc
down to $10^{14} \; \mathrm{cm}^{-2}$ out to the virial
radius \citep{2014ApJ...792....8W}, corresponding to number
densities of around $10^{-2}$ to
$10^{-3} \; \mathrm{cm}^{-3}$
\citep{2019MNRAS.484.2257Z,2015MNRAS.446...18C}. In such an
environment, Jeans' instabilities alone are insufficient in
forming the cool phase of the CGM. Turbulence, evidenced by
observations of broad line-widths
\citep{2016ApJ...833...54W,2012ApJ...750...67R}, plays a
pivotal role in the gas dynamics of the CGM. A key driver in
nonvirialized halos with high accretion rates
($> 10 \; M_\odot \; \mathrm{yr}^{-1}$ for $10^{12}M_\odot$
haloes), as mentioned previously, are supersonic cool
accretion flows which can stir turbulence via localized
thermodynamics such as turbulent mixing between hot and cold
phases and gas entrainment onto supersonic cold streams
\citep{2019MNRAS.487..737J,2023MNRAS.520.2148Y}. Internal
feedback processes from supernovae and/or active galactic
nuclei (AGN) within the galaxy, are also capable of driving
turbulence in the CGM via large-scale outflows
\citep{2017MNRAS.466.3810F}. These feedback-induced
turbulence motions in turn significantly impact the
accretion of gas back onto the ISM or into the AGN,
rendering it chaotic and asymmetric
\citep{2013MNRAS.432.3401G}. They can also “stimulate” the
clumping and eventual precipitation/accretion of CGM gas
back onto the ISM in regimes where $t_\mathrm{cool}$ exceeds
$t_\mathrm{ff}$
\citep{2017ApJ...845...80V,2018ApJ...868..102V}. Within a
cold cloud, turbulence may fragment the cloud into smaller
"droplets", exponentially increasing the surface area and
growth rate of cool clumps \citep{2022MNRAS.511..859G},
which may explain the observational evidence for cold
streams in the CGM. Additionally, turbulent winds can damp
fluid instabilities in cool clouds in the presence of hot
winds, stabilizing the cloud against destruction
\citep{2020MNRAS.499.4261S}, although compared to radiative
cooling processes this only has a significant impact in
highly supersonic flow \citep{2020MNRAS.492.1841L}.

However, turbulence in such diffuse astrophysical media is
not necessarily a continuous phenomenon, particularly in hot
virialized haloes where accretion is subsonic and
quasi-spherically symmetric
\citep{2023ARA&A..61..131F}. Internal sources such as
feedback are tied to episodic starburst and AGN activity
from the disc
\citep{2014ApJ...794..156R,2015ApJ...812...83N,2013ApJ...768...18B},
and environmental sources are tied to episodic mergers or
ram pressure on infall into a group or cluster. Hence, this
necessitates the study of not merely steady-state turbulence
driving, but also the dissipation of turbulence during
quiescent periods. Using magnetohydrodynamic simulations of
turbulence driving on molecular cloud scales with an
isothermal equation of state, \citet{1998ApJ...508L..99S}
found saturation timescales to be independent of magnetic
field strength, and energy dissipation timescales to be
inversely correlated with magnetic field strength once
turbulence was turned off. Magnetic fields provide magnetic
pressure, which plays an important role in the gravitational
stability of GMCs \citep{2001ApJ...546..980O}, although the
role it plays in CGM cloud stability is uncertain and
somewhat of an open question
\citep{2023ARA&A..61..131F}. While
\citet{1998ApJ...508L..99S} considered variable magnetic
field strength (damping strength) across multiple runs, they
fixed their turbulence driving strength.

In this study, using hydrodynamic simulations, we
investigate the effects of a generalized source of
turbulence in a circumgalactic environment, though it can be
extended to broader environments such as the intracluster
medium or diffuse phases of the ISM. This work applies and
builds on the methodology of \citet{1998ApJ...508L..99S} to
such environments, with variable turbulence driving
strengths, an adiabatic equation of state and a standard
collisional ionization equilibrium (CIE) cooling curve
\citep{2012ApJS..202...13G}. We will elaborate on the
methodological differences from \citet{1998ApJ...508L..99S}
in Section \ref{sec:methods}.

\section{Methods}
\label{sec:methods}

\subsection{Hydrodynamics and Thermodynamics }

Studying turbulence numerically requires solving a
consistent set of hydrodynamic equations.  We employ the
hetrogeneous hydrodynamic code \kratos (Wang et al. in
prep), integrating the following Euler equations for an
ideal gas:
\begin{equation}
  \label{eq:method-base-eqs}
  \begin{split}
    \dfrac{\partial \rho}{\partial t} + \nabla \cdot (\rho
    \mathbf{v})
    & = 0\ ,\\
    \dfrac{\partial (\rho \mathbf{v})}{\partial t}
    + \nabla \cdot (\rho \mathbf{v} \mathbf{v} +
    p\mathbf{I})
    & = -\rho \nabla\Phi ,\\
    \dfrac{\partial \varepsilon}{\partial t}
    + \nabla \cdot [\mathbf{v}(\varepsilon + p) ]
    & = -\rho \mathbf{v}\cdot \nabla\Phi + S \ .
  \end{split}
\end{equation}
Here, $\rho$, $\mathbf{v}$, $p$ and $\epsilon$ are the mass
density, velocity, gas pressure, and total energy density,
respectively. $\Phi$ is the gravitational potential,
$\mathbf{I}$ is the identity tensor, and $S$
is the source term which captures additional thermodynamics beyond adiabatic 
compression and expansion. The
hydrodynamic solver employs the slope-limited piecewise
linear method (PLM) reconstruction scheme, the HLLC approximate Riemann
solver, and the second order Runge-Kutta (RK2) time
integrator. The Jeans length under typical CGM physical
parameters approximately reads,
\begin{equation}
  \label{eq:method-jeans}
  l_J \sim  10^4~\pc \times
  \left( \dfrac{T}{10^4~\K} \right)^{1/2}
  \left( \dfrac{\rho}{10^{-2}~m_p~\cm^{-3}} \right)^{-1/2}\ .
\end{equation}
Such scales are considerably greater than the box size, let alone the
typical cloud sizes, within our turbulent CGM simulations, making
$\Phi \rightarrow 0$ a safe approximation in this study.

\begin{figure}
    \centering
    \incgraph{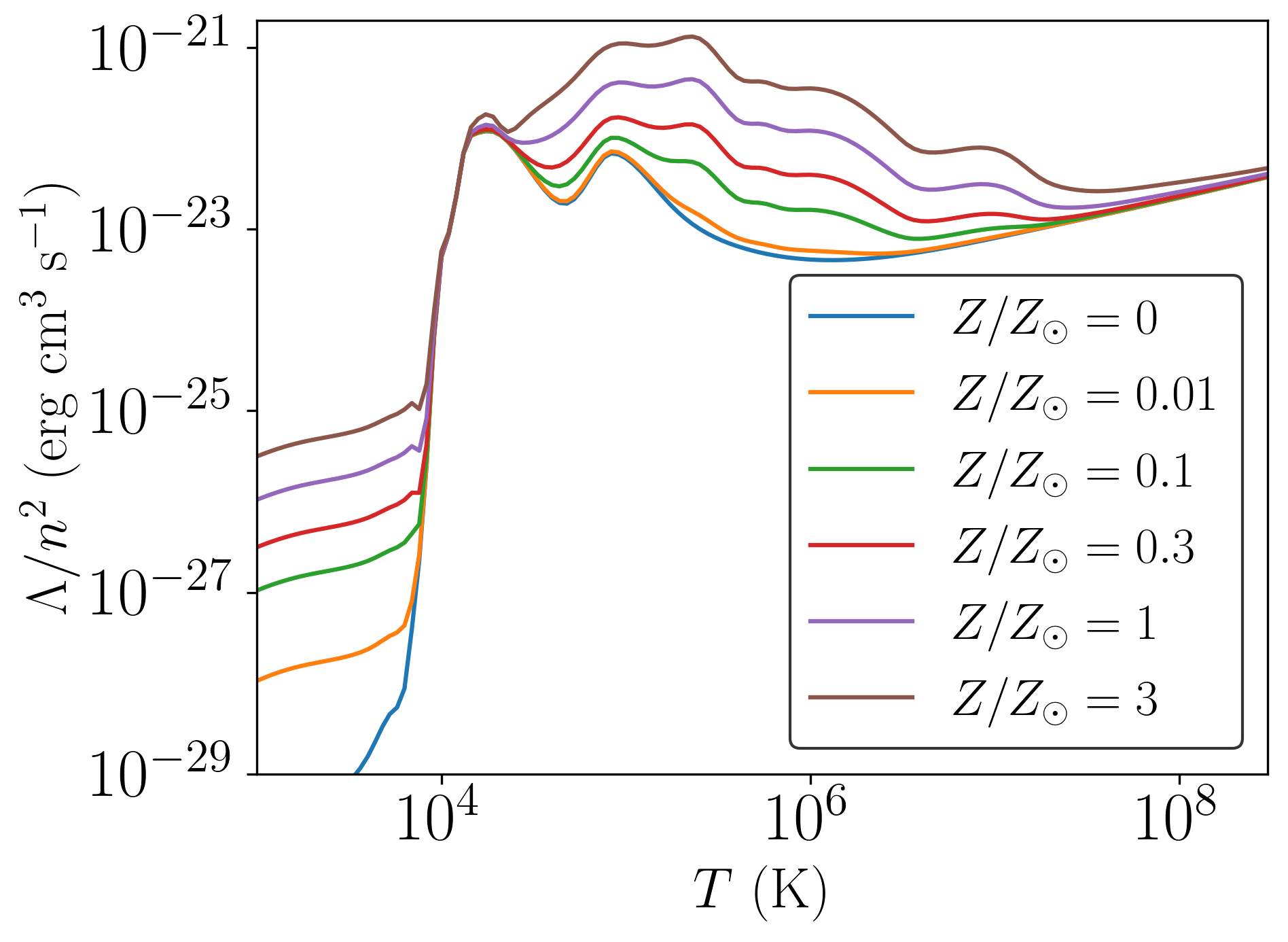}
    \caption{The cooling curves used in the simulations,
      adopting the data provided by
      \citet{2012ApJS..202...13G}. The interpolated and
      extrapolated curves are computed as
      $\Lambda(Z/Z_\odot)/n^2 = \Lambda(0)/n^2 +
      (Z/Z_\odot)(\Lambda(1)/n^2 - \Lambda(0)/n^2)$.}
    \label{fig:cool}
\end{figure}

Within each sub-step in the RK2 cycles, we integrate the impact
of $S$ in every zone via a semi-implicit method.  The cooling
rate, represented by $S$ in eq.~\eqref{eq:method-base-eqs},
is based on the standard table from
\citet{2012ApJS..202...13G}. It covers the temperature
range $10^3-10^8~{\rm K}$ and varying metallicities (see 
Figure \ref{fig:cool}). For typical CGM gases with
density $\rho \lesssim 10^{-2}~m_p~\cm^{-3}$, the column
density required for the cooling photons to escape in typical CGM
scenarios is estimated by $N\lesssim 10^{19}~\cm^{-2}$ per
kpc along the escape path. For dense clouds with $\sim 10^2\times $ the mean density, the typical
sizes ($\sim 10^1~\pc$) are sufficiently small to allow
cooling photons to escape. These are sufficient to guarantee
the optical-thin condition for the standard cooling table.

\subsection{Turbulence Driving}

This study focuses on the local dissipative
behaviors of various phases of the CGM, hence the turbulence energy cascade
is emulated via kinetic energy injection. We follow an approach similar to the one outlined
in \citet{1999ApJ...524..169M}. During each timestep, the
simulation domain is subject to a uniform perturbation via
an acceleration vector aligned with a randomly selected
direction vector $\hat{a}$. Denoting the amplitude of
perturbation as $A$, the turbulence energy injection rate
per by unit mass $\dot{\epsilon}_i$ (which will be referred to as turbulence driving strength) can be evaluated as,
\begin{equation}
  \label{eq:turb-inj-norm}
  \dot{\epsilon}_i = A \left\{ [\langle\rho \rangle L^3]^{-1}
    \Delta t \  \sum_{\mathbf{i}} \delta V_{\mathbf{i}}\;
    \rho_{\mathbf{i}} \mathbf{v}_\mathbf{i} \cdot \hat{a}\right\}\ ,
\end{equation}
where $L$ is the box length, $\langle \rho \rangle$ is the
mean mass density throughout the domain, and $\Delta t$ is
the current timestep. The summation goes through all cells,
where $\mathbf{i}$ is the cell index, and $\delta V_\mathbf{i}$,
$\rho_\mathbf{i}$ and $\mathbf{v}_\mathbf{i}$ are the cell
volume, mass density, and velocity vector for the
$\mathbf{i}$-th cell.  In practice, we establish a constant
$\dot{\epsilon}_i$ as a key model parameter, which remains
fixed throughout the simulation's progression up until turbulence turnoff. For each
timestep, we utilize the enclosed summation in
eq.~\eqref{eq:turb-inj-norm} to normalize the amplitude
$A$. Subsequently, we inject kinetic energy into the
system by updating the velocity vector to
$\mathbf{v}_\mathbf{i}' = \mathbf{v}_\mathbf{i} + A\hat{a}
\Delta t$, and update the energy density accordingly.

\subsection{Simulation Setup}
\label{sec:method-setup}

 Focusing on the local structures and characteristics
  of turbulence compression and subsequent dissipation
  processes, these simulations are conducted on
  $L = 256~\pc$ cubic domains with periodic boundary
  conditions. Various simulations are carried out to cover
  the parameter space, with 6 different metallicity values
  $Z/Z_\odot \in \{0, 0.01, 0.1, 0.3, 1, 3 \}$, 6 different
  average gas density values
  $\langle \rho \rangle / (m_p~\cm^{-3}) \in \{10^{-4},
  10^{-3}, 10^{-2}, 10^{-1}, 10^0, 10^1\}$, and 7 different
  turbulence energy injection rate values
  $\dot{\epsilon}_i / (\erg~\s^{-1}~\g^{-1}) \in \{0.01,
  0.1, 0.3, 1, 3, 10, 100\} $. In total, there are 252
  simulations. Selection of these parameters extend beyond
the density and metallicity range typical of the CGM to
account for denser, metal enriched phases captured in cool
clouds, though we note that the Jeans length remains larger
than our box even for our highest density runs.

Methodologically, setting aside the different astrophysical
medium of interest and our exclusion of magnetic fields, a
key difference between this work and
\citet{1998ApJ...508L..99S} is our usage of an adiabatic
equation of state with the inclusion of thermal energy,
adiabatic heating and a radiative cooling function. This is
necessary given the differences between CGM and GMC
environments: the CGM is orders of magnitude hotter and
reside in efficient cooling temperature regimes, and is
subject to much more violent turbulence driving
events. Without artificial viscosity, energy dissipation
happens entirely via the radiative cooling function, with
kinetic energy dissipating via the energy cascade into
thermal energy.

The wide coverage over the
multi-dimensional parameter space limits the 
viable resolution of each run due to computational and 
storage limits. We adopt a relatively low $128^3$
resolution for each run, resulting in a spatial resolution
of $2~\pc$. Such resolution is sufficient to resolve
compressive shocks that form relatively dense
gas clumps with sizes $\sim 10~\pc$ (see Section \ref{sec:cloudcrush}).  For conditions
matching more closely to those of the dense phase of the
CGM, namely $Z/Z_\odot \in \{0.1, 0.3\}$ and
$\rho/(m_p~\cm^{-3}) \in \{10^{-2}, 10^{-1}\}$, we run an
additional set of higher resolution $256^3$ runs with the
exact same setup otherwise. The denser and higher
metallicity phases are chosen given their efficient cooling
regimes, which allows the turbulence to remain supersonic
for higher values of $\dot{\epsilon}_i$, and thus more
analyzable high resolution runs.

The initial conditions of each run are set so that
the initial gas temperature is uniformly
$T = 3\times 10^4~\K$. 

The initial velocity field is randomized, with each
component of each cell's velocity having a random value
between $\pm 1~\km~\s^{-1}$,
representing a homogenous gas with initial subsonic
motions. The choice of initial velocity field is arbitrary
as long as the velocities are subsonic, since all runs will
undergo turbulence driving until the gas reaches saturated
steady-state turbulence. An initial non-zero velocity field
is necessary, since given a periodic box and entirely
uniform perturbations, theoretically no substructures would
form, and any resulting substructures that do form are the
result of numerical errors and approximations in the flux
advection and reconstruction schemes. The randomized
subsonic initial motions guarantee a physical basis for any
resulting turbulence driven substructures.

Each run is fixed to 210000 cycles.  Turbulence
  driving is turned on for the first 200000 cycles to
  guarantee turbulence saturation and a 
  turbulent quasi-steady states. Turbulence is turned off for the last 10000
  cycles to study the dissipation of turbulences. This scheme of
  fixing the number of cycles is related to the
  Courant-Friedrichs-Lewy (CFL) conditions, which limits the
  timestep of each cycle $\Delta t$ by
\begin{equation}
  \label{eq:method-cfl}
  \Delta t = C \times \min_{\mathbf{i}}
  \bigg\{ \dfrac{\Delta x}
    {|\mathbf{v}_{\mathbf{i}}| + c_{s,\mathbf{i}}}
  \bigg\}\ .
\end{equation}
In this work we choose $C = 0.3$. As one can easily deduce, each cycle with timestep
constrained by eq.~\eqref{eq:method-cfl} represents a
fraction of the fluid crossing time (in case of high Mach
numbers) or sound crossing time (in case of low Mach
numbers). Therefore, a fixed number of cycles naturally
covers a similar number of fluid or sound crossing
timescales, providing sufficient temporal resolution for analysis 
while minimizing data storage requirements.

\section{Results}
\label{sec:results}

Unless specified otherwise, plots showing a sample of data
use the $128^3$ runs, while plots showing single
representative samples use the $256^3$ runs. Plots showing
$256^3$ runs will have this fact relayed in figure captions,
and we will discuss resolution convergence in Appendix
\ref{sec:appendix:res}.

\subsection{Subsonic and Supersonic Dissipation} \label{sec:results:twoepoch}
\begin{figure}[h!]
    \centering
    \incgraph{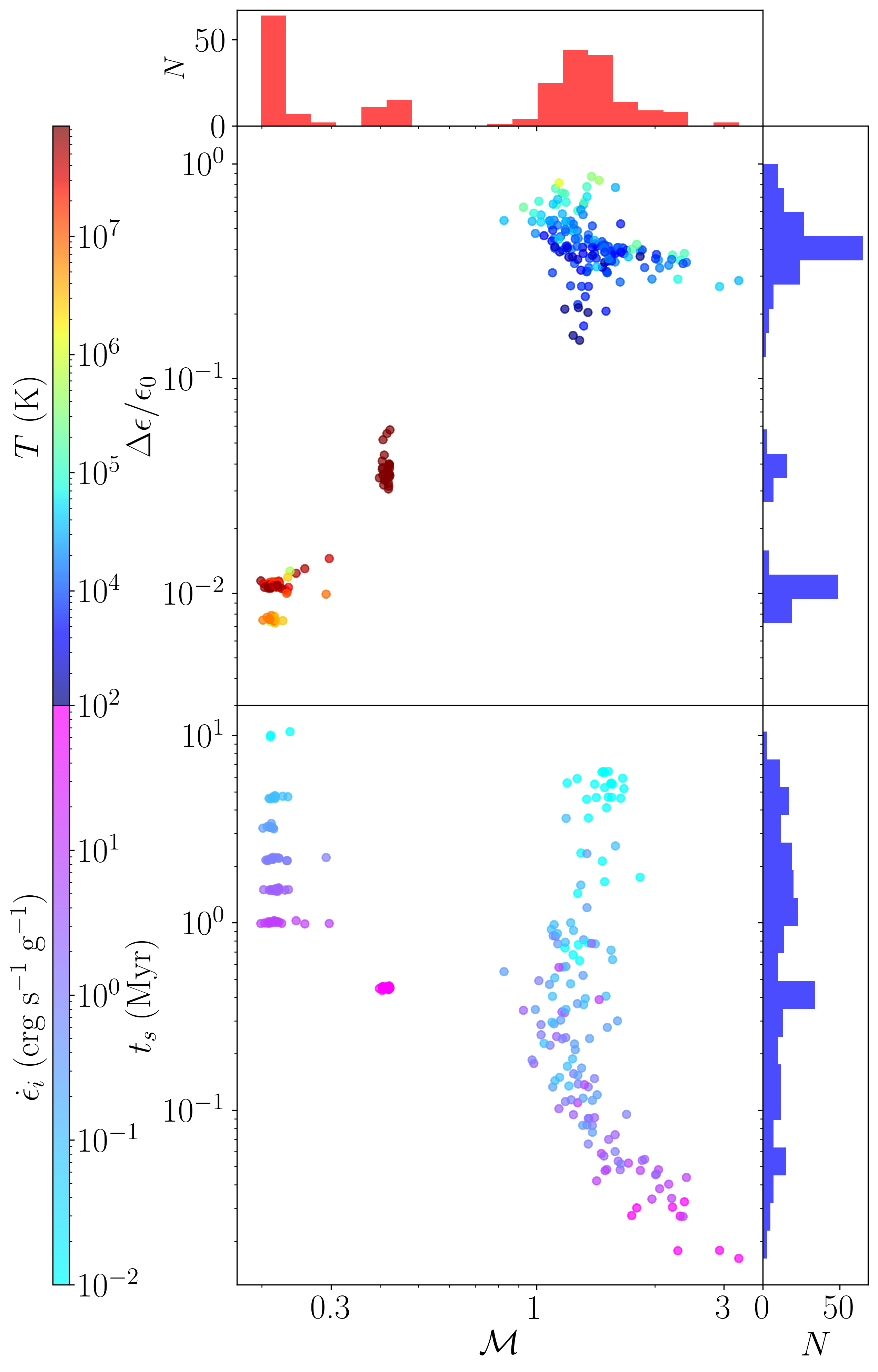}
    \caption{Two scatter plots showing the mass averaged mach number ($\mathcal{M})$ vs the fractional energy decay at three saturation times ($\Delta \epsilon / \epsilon_0$)  on top and the kinetic energy saturation time ($t_s$) on the bottom, with each point representing a run. $\mathcal{M}$ is defined as $v_\mathrm{avg} / c_{s,0}$, where $v_\mathrm{avg}$ is the mass averaged velocity across all cells and the initial sound speed $c_{s,0}$ is computed from the mass-averaged temperature of the entire box. $t_s$ represents the kinetic energy saturation time defined as the ratio of the kinetic specific energy to the turbulence driving strength $t_s = \epsilon_k / \dot{\epsilon}_i$. The temperatures on the top plot represent initial temperatures at turbulence turnoff.}
    \label{eneloss_mach}
\end{figure}

\begin{figure*}
    \centering
    \incgraph{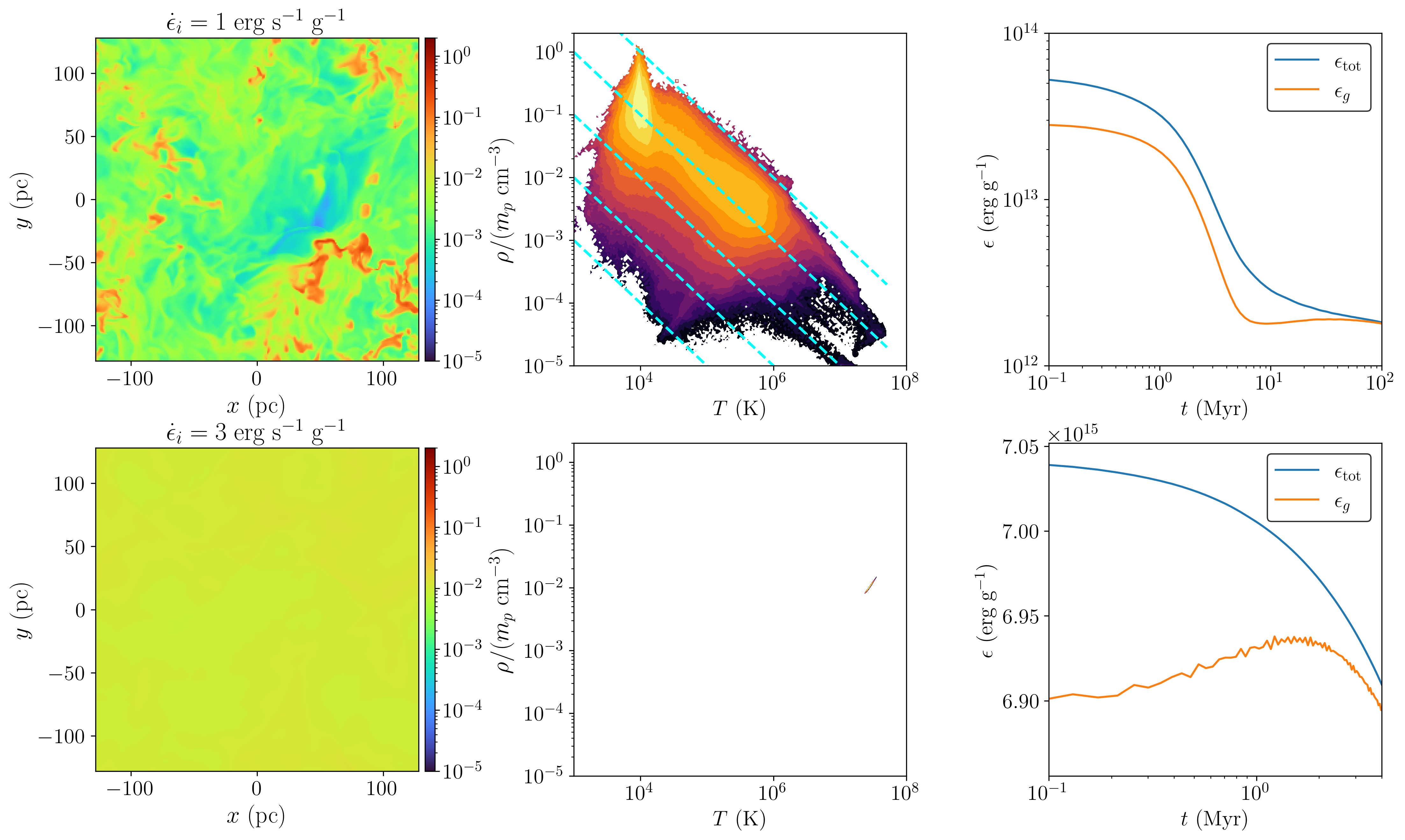}
    \caption{A comparison between two $256^3$ resolution runs where $Z/Z_\odot = 0.3$ and $n = 10^{-2} \; \mathrm{cm}^{-3}$, with the top and bottom rows representing $\mathrm{\dot{\epsilon}_i} = 1 \; \mathrm{erg} \; \mathrm{s}^{-1} \; \mathrm{g}^{-1}$ and $\mathrm{\dot{\epsilon}_i} = 3 \; \mathrm{erg} \; \mathrm{s}^{-1} \; \mathrm{g}^{-1}$ respectively. The left column shows density slice plots at $z=0$, the middle column the mass-weighted density-temperature phase plots (contoured 2D histogram), and the right column the specific total and thermal energies and the mass-averaged temperature over time.  The dotted blue lines in the top middle phase plot represent lines of constant pressure.}
    \label{phasecompare}
\end{figure*}

A bimodality in the initial turbulent state and the energy dissipation is seen in Figure \ref{eneloss_mach}, with a distinct supersonic and subsonic turbulence regime. In the subsonic case, the initial averaged mach numbers ($\mathcal{M}$) are less than $1$, the initial temperatures are above $10^6$ K, and little dissipation is present within $3t_s$. Meanwhile in the supersonic case, the $\mathcal{M}$ values are of order unity and above, the initial temperatures are below $10^5$ K, and significant energy dissipation of $10-80\%$ within three kinetic saturation timescales is present within $3t_s$. There is a temperature gradient across the supersonic runs, where hotter steady-state turbulence leads to increased energy dissipation. Among the subsonic runs, the energy loss hovers around $1\%$, but runs with $\dot{\epsilon}_i = 100 \; \mathrm{erg} \; \mathrm{s}^{-1} \; \mathrm{g}^{-1}$ and initial temperatures of around $10^8$ K lose around $3-6\%$ of their initial energy within $3t_s$, with initial $\mathcal{M}$ higher than the other subsonic runs. The distribution of fractional energy losses almost seems multimodal, with two distinct subsonic regimes and one distinct supersonic regime. Across runs with fixed initial conditions (same $n$ and $Z/Z_\odot$), increasing $\dot{\epsilon}_i$ leads to an eventual transition point from supersonic to subsonic turbulence (see Figure \ref{phasecompare}, though this offset in both $\mathcal{M}$ and $\Delta \epsilon/\epsilon_0$ between the subsonic $\dot{\epsilon}_i = 100 \; \mathrm{erg} \; \mathrm{s}^{-1} \; \mathrm{g}^{-1}$ and other subsonic runs may imply that with even stronger turbulence driving there is another supersonic regime. The transition occurs when thermal dissipation via radiative cooling is unable to reach an equilibrium with energy injection via turbulence driving and energy cascade until the gas temperature exceeds $10^7$ K and enters the bremsstrahlung regime. Regions along the cooling curve (see Figure \ref{fig:cool}) where $\partial \Lambda/\partial T < 0$ are unstable during turbulence driving, hence giving rise to a temperature (and by proxy the subsonic/supersonic turbulence) bimodality.

Examining the distribution of kinetic saturation times, which is defined as $t_s = \epsilon_k / \dot{\epsilon}_i$ during steady state turbulence,  we observe it to be smooth and continuous among the supersonic runs, with timescales ranging from $3\times10^{-2} - 10^1 \; \mathrm{Myr}$. As with temperature, we see a $\dot{\epsilon}_i$ gradient where stronger turbulence driving leads to lower $t_s$. $\mathcal{M}$ also shows a peculiar trend where it scales positively with $\dot{\epsilon}_i$ for $t_s > 3 \times 10^{-1} \; \mathrm{Myr}$, but negatively with $\dot{\epsilon}_i$ for $t_s < 3 \times 10^{-1} \; \mathrm{Myr}$ The subsonic runs appear clustered, with distinct vertical offsets in $t_s$ congruent to distinct values of $\dot{\epsilon}_i$, which is far different from the continuous distribution of $t_s$ among the supersonic runs. This would suggest in the subsonic regime, $t_s$ depends primarily on $\dot{\epsilon}_i$, with weaker dependancies (if any) on $n$ or $Z$.

The physical distinction between subsonic and supersonic dissipation becomes clearer in Figure \ref{phasecompare}, where we examine the steady state of two particular $256^3$ resolution runs in more detail. On the top where $\mathrm{\dot{\epsilon}_i} = 1 \; \mathrm{erg} \; \mathrm{s}^{-1} \; \mathrm{g}^{-1}$, we observe large variations in $\rho$ and $T$ across many orders of magnitude, and distinct overdense clumps in the gas. Meanwhile on the bottom, where $\mathrm{\dot{\epsilon}_i} = 3 \; \mathrm{erg} \; \mathrm{s}^{-1} \; \mathrm{g}^{-1}$, we observe a near uniform gas with perturbations of order unity. Subsonic motions in the gas are unable to produce shocks, which is a prerequisite for substructure formation in diffuse, non self-gravitating regimes. The right column paints a clearer picture of the different regimes of dissipation seen in Figure \ref{eneloss_mach}. Supersonic turbulence sees two distinct epochs of dissipation, characterized by an initial rapid dissipation epoch followed by a slow dissipation epoch. Subsonic turbulence on the other hand, sees only a slow dissipation epoch, cotemporal with an increase in thermal energy. This increase can also be seen in thermal energy of the supersonic run, and represents the cascading rate of kinetic energy into thermal energy via numerical hydrodynamic processes exceeding the radiative cooling rate. The physical nature behind the two dissipation epochs and the individual dissipative behaviours of thermal and kinetic energy will be examined more closely in Section \ref{sec:results:thermalvskinetic}.

\begin{figure} 
    \centering
    \incgraph{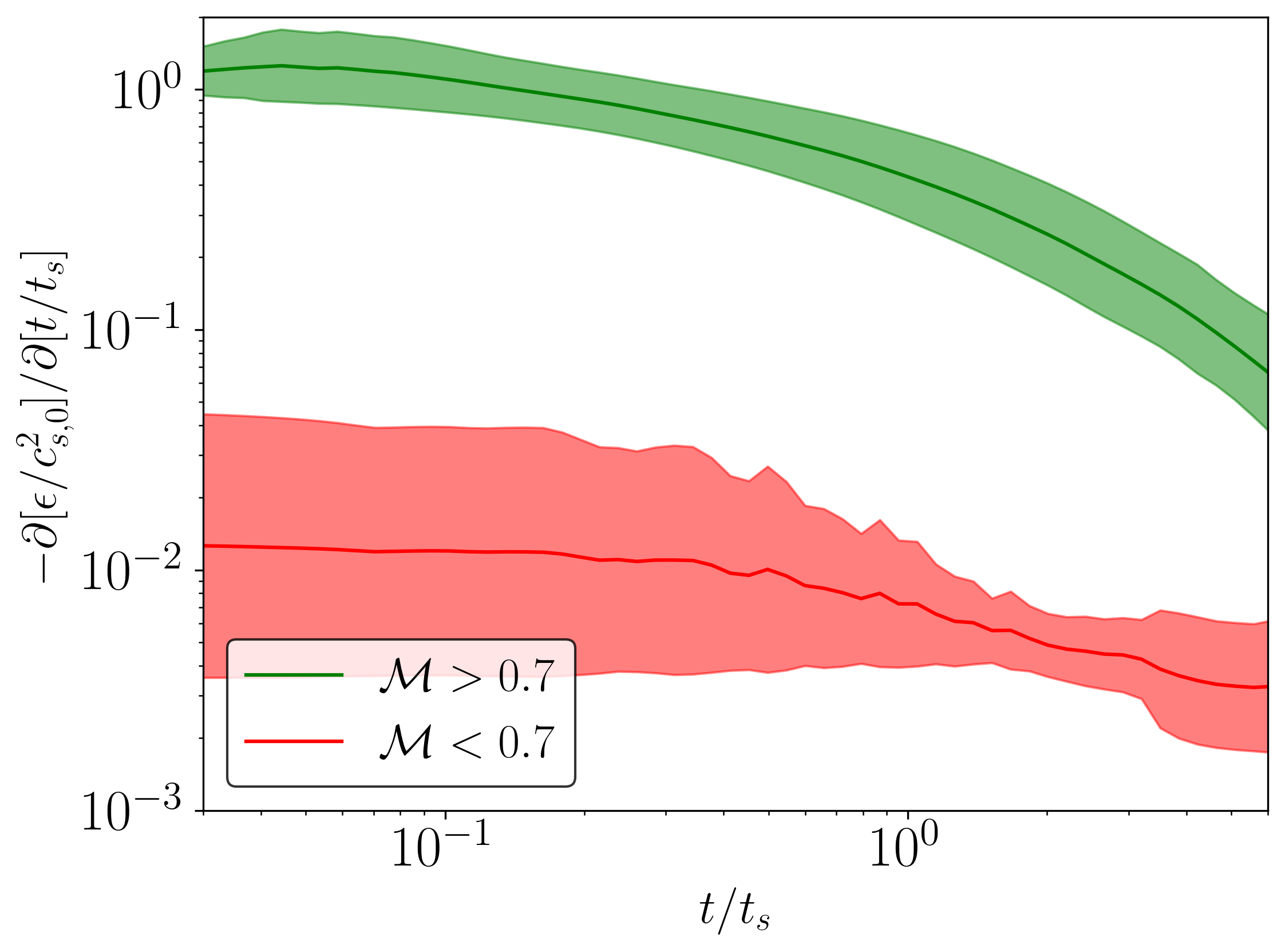}
    \caption{The dissipation rate of specific total energy, normalized by initial sound speed $c_{s,0}^2$, over time for supersonic (green) and subsonic runs (red). $\mathcal{M}$ follows the same definition as the one in Figure \ref{eneloss_mach}, which is $v_\mathrm{avg}/c_{s,0}$, and $t_s$ represent the kinetic energy saturation time. The shaded error regions represent 1$\sigma$.}
    \label{dedt}
\end{figure}

This distinction between the dissipation rates of supersonic and subsonic turbulence can be seen in Figure \ref{dedt}. Here the energy is normalized by the initial sound speed at turbulence turnoff $c_{s,0}^2$ consistent with \citet{1998ApJ...508L..99S}. The normalized dissipation rates of the supersonic runs as a whole are a full dex higher than those of the subsonic runs up to a few $t_s$, beyond which the rates begin to converge. The convergence is indicative of the dissipation of supersonic motions and dispersal of substructures as the box becomes homogenized, and is discussed more in
depth in the following sections. Figure \ref{dedt} encapsulates most of the entire range in $t_s$ spanned by the subsonic runs, with the supersonic runs spanning a larger range up to hundreds and thousands of $t_s$. As mentioned earlier, this is the result of the shorter timesteps of the subsonic runs.

\subsection{Thermal vs Kinetic Supersonic Dissipation} \label{sec:results:thermalvskinetic}

\begin{figure}[h!]
    \centering
    \incgraph{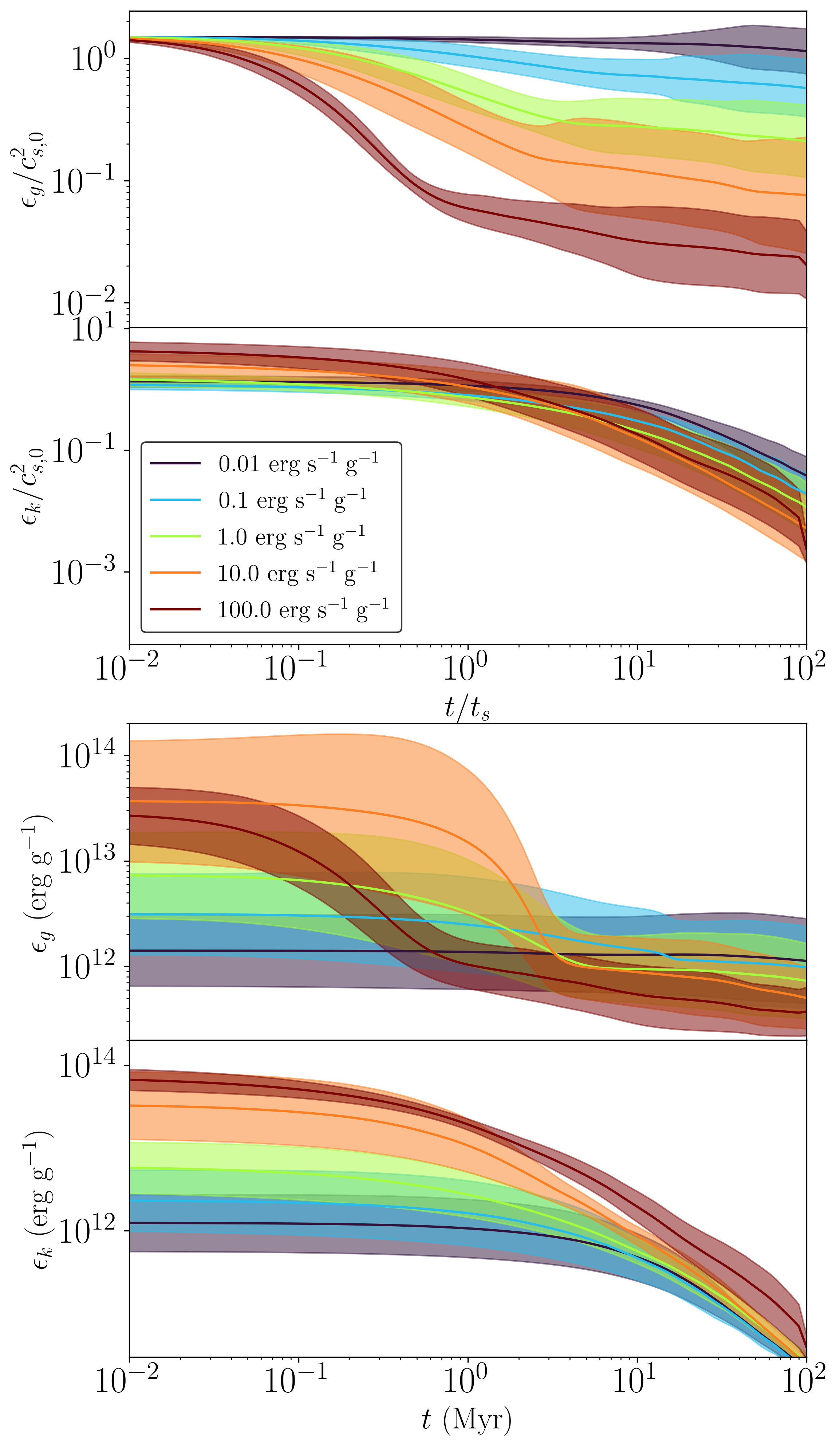}
    \caption{Normalized (top two) and unnormalized (bottom two) energy dissipation curves over time. Specific energy is normalized by initial sound speed $c_{s,0}^2$, and time is normalized by kinetic saturation time $t_s$. The shaded regions represent 1$\sigma$ error regions and each coloured curve represents different values of $\mathrm{\dot{\epsilon}_i}$. The first and third plots show thermal energy, and the second and fourth plots show kinetic energy.}
    \label{enetime}
\end{figure}

Fixing our attention towards the supersonic runs, Figure \ref{enetime} shows the time evolution of specific thermal ($\epsilon_g$) and kinetic ($\epsilon_k$) energy binned in $\dot{\epsilon}_i$, with the top two plots normalized by initial sound speed $c_{s,0}^2$ at turbulence turnoff and $t_s$. It is immediately clear that $\epsilon_g$ and $\epsilon_k$ dissipate very differently on very different timescales. In the normalized plots, there are significant variations in thermal dissipative behaviour with respect to $\dot{\epsilon}_i$. Beyond $10t_s$, 1 dex increases in $\dot{\epsilon}_i$ result in an approximate 0.5 dex drop in normalized thermal energy. Meanwhile, kinetic dissipation curves are more or less within each others' error regions, although there is some weak dependancy on $\dot{\epsilon}_i$, with stronger turbulence driving leading to slightly shorter dissipation timescales relative to $t_s$.

\begin{figure}
  \centering
  \incgraph{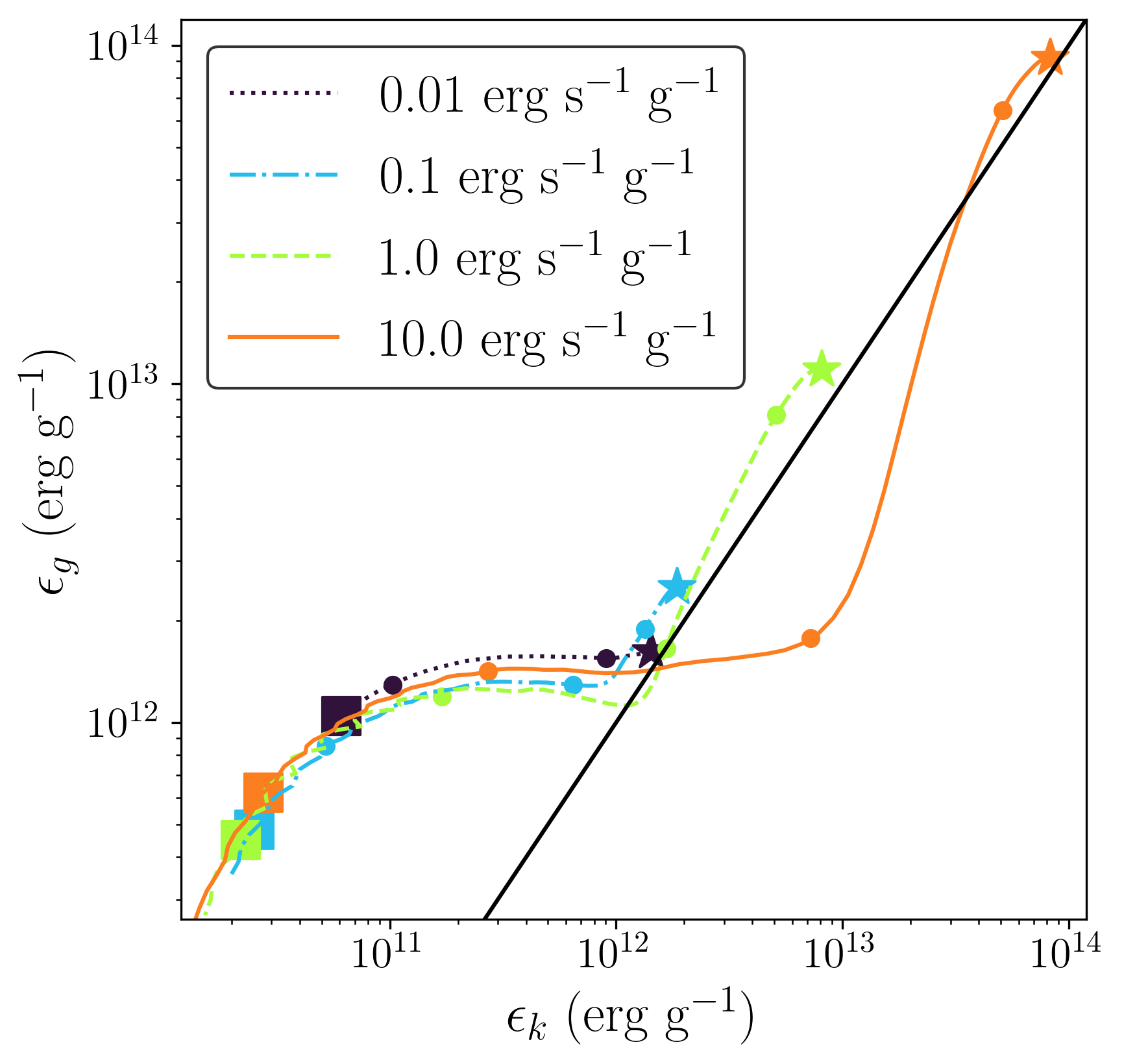}
    \caption{Phase plots showing the time tracks of thermal versus kinetic energy for $256^3$ resolution runs corresponding to $Z/Z_\odot = 0.3$ and $n = 0.1 \; \mathrm{cm}^{-3}$. Stars denote the turbulence driving turn off time where each curve begins, circular dots denote $t_s$, $10t_s$ and $100t_s$ as one moves from the star and along the curve, and the squares denote $t = 100 \; \mathrm{Myr}$. The black line represents the $x=y$ line.}
    \label{enephasebig}
\end{figure}

\begin{figure}
  \centering
  \incgraph{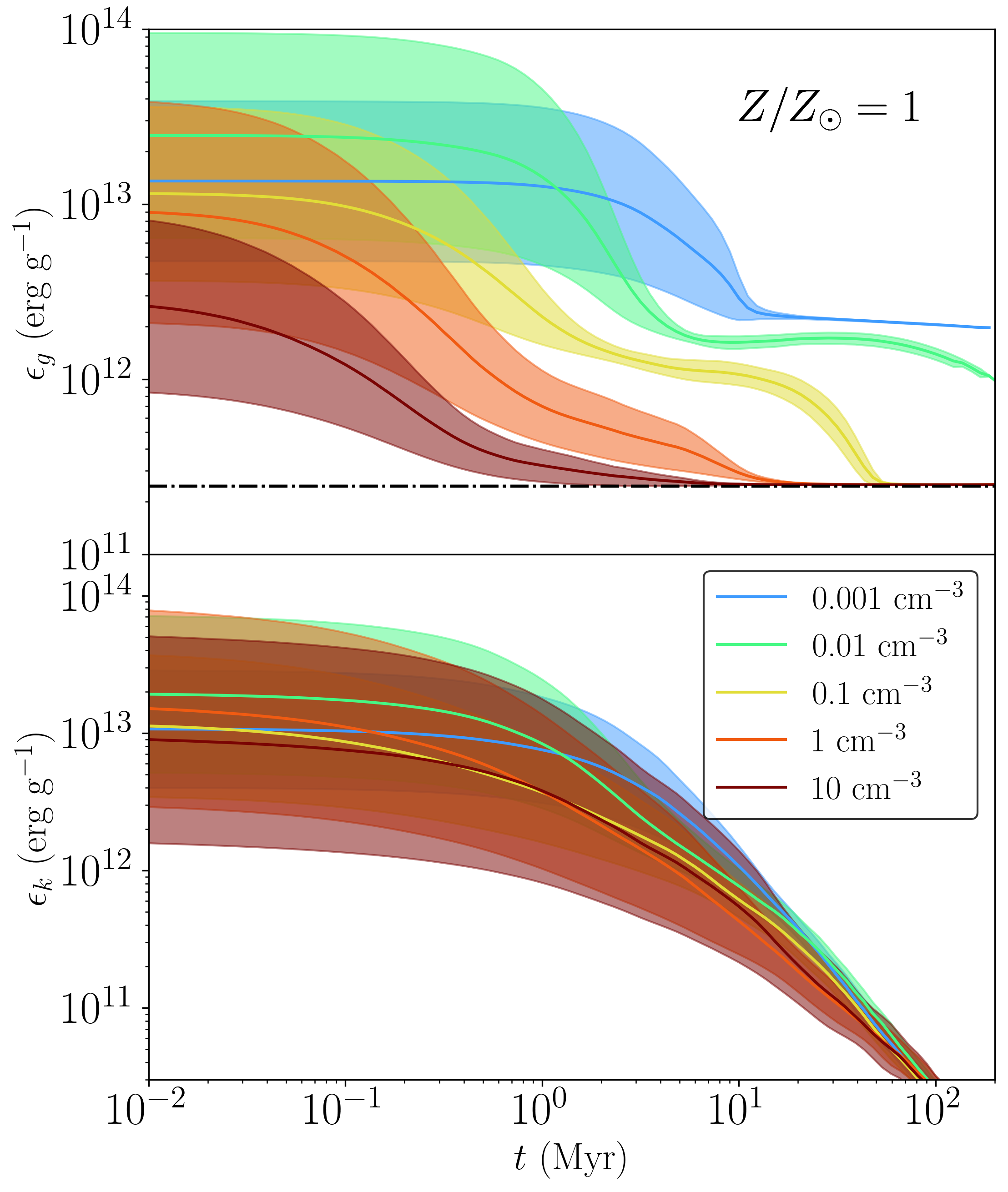}
    \caption{A similar plot as Figure \ref{enetime} but with curves binned in $n$ as opposed to $\dot{\epsilon}_i$. The shaded regions represent 1$\sigma$. The thermal energy (top) plot shows only $Z/Z_\odot = 1$ runs with the shaded regions only representing variations in $\dot{\epsilon}_i$, while the kinetic energy (bottom) plot shows all metallicity runs with the shaded regions representing variations in both $Z$ and $\dot{\epsilon}_i$. The black dotted line represents the specific thermal energy floor corresponding to the $1000$ K temperature floor.}
    \label{enetimenonorm}
\end{figure}

Thermal dissipation is characterized by the initial rapid dissipation epoch as previously seen in Figure \ref{phasecompare}, on timescales of order $0.1t_s$. The separation between different binned values of $\dot{\epsilon}_i$ is somewhat orderly in the normalized plot, manifesting themselves as a "branch-off" from their universal initial values of $c_{s,0}^2/(\gamma - 1)$ starting at $0.01t_s$. Higher $\dot{\epsilon}_i$ sees higher amounts of normalized thermal dissipation during the rapid epoch, before all runs eventually see dissipation rates decreasing significantly and their curves flattening. The bottom unnormalized thermal dissipation plot reveals a degree of non-linearity in the starting times of the rapid dissipation epoch with the $\dot{\epsilon}_i = 10 \; \mathrm{erg} \; \mathrm{s}^{-1} \; \mathrm{g}^{-1}$ beginning at later times compared to the other runs. The unnormalized plot also reveals a range of energy plateaus shared across all runs afterwards, representing an inefficient cooling regime once the gas has reached $10^4$ K, and will be examined in more detail in Figures \ref{enephasebig} and \ref{enetimenonorm}. Kinetic dissipation is characterized by a steady exponential drop over time. Unlike thermal dissipation, kinetic dissipation is far more weakly coupled to $\dot{\epsilon}_i$, despite the kinetic energy during steady state turbulence having a strong dependency on $\dot{\epsilon}_i$ as seen in the initial values of the kinetic dissipation plots, both normalized and unnormalized. In the unnormalized plot, despite the gradient of initial $\epsilon_k$, the dissipation curves all converge after $100$ Myr.

\begin{figure*}
    \centering
    \incgraph{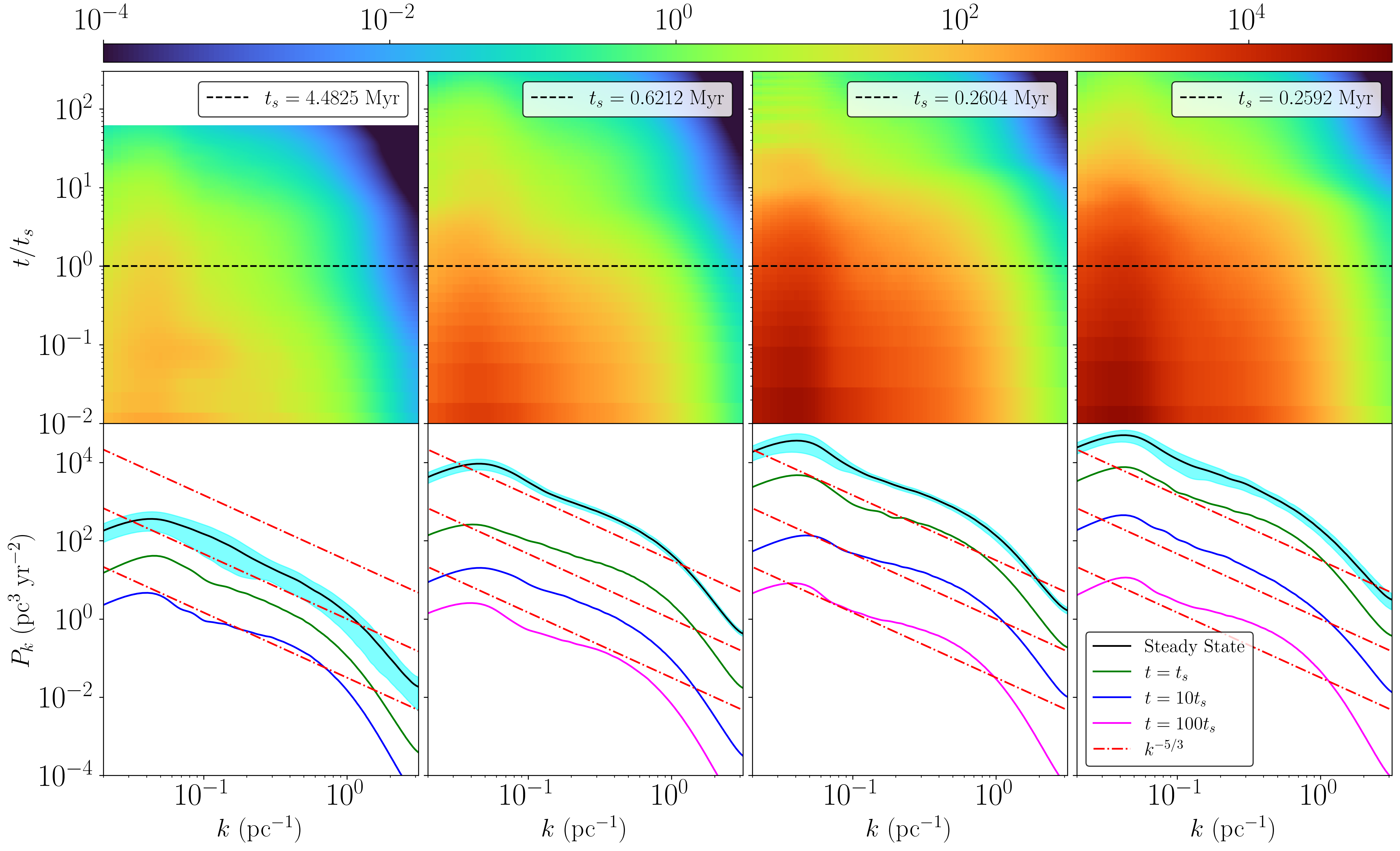}
    \caption{The turbulence power spectra of the four high resolution $256^3$ runs shown in Figure \ref{enephasebig}, where $Z/Z_\odot = 0.3$ and $n = 0.1 \; \mathrm{cm}^{-3}$, and $\dot{\epsilon}_i$ being 0.01, 0.1, 1.0 and 10.0 $\mathrm{erg} \; \mathrm{s}^{-1} \; \mathrm{g}^{-1}$ from left to right. The bottom row plots the turbulence power spectra during steady state turbulence, after $t_s$, $10t_s$ and $100t_s$, as well as $k^{-5/3}$ reference lines representing the inertial regime of the kolmogorov spectrum. The cyan coloured region represents $1\sigma$ across the last twenty outputs of the steady state turbulence epoch. The top row plots the power spectra from turbulence turn off up to $100t_s$, with the colour denoting $\epsilon_k$. The value of $t_s$ is also shown across each of the four runs. Both the top and bottom row plots are smoothed using degree-5 polynomial splines for visualization purposes. The leftmost $\dot{\epsilon}_i = 0.01 \; \mathrm{erg} \; \mathrm{s}^{-1} \; \mathrm{g}^{-1}$ column does not show a $100t_s$ power spectrum since the simulation run ended before $100t_s$.}
    \label{powspecspline}
\end{figure*}

We show the time tracks of a few higher resolution runs comparing thermal and kinetic energy directly in Figure \ref{enephasebig}. We observe rapid dissipation in thermal energy within the first $10t_s$ for higher values of $\dot{\epsilon}_i$, and relatively consistent exponential kinetic dissipation rate with no plateaus, consistent with our observations of Figure \ref{enetime}. The latter can be further evidenced by the approximately even horizontal spacing between circles (most pronounced for $\dot{\epsilon}_i = 10 \; \mathrm{erg} \; \mathrm{s}^{-1} \; \mathrm{g}^{-1}$), with $\epsilon_k$ being a viable proxy for time. We also confirm the convergence of kinetic and thermal energy as seen in the unnormalized bottom plots of Figure \ref{enetime} - regardless of initial energy or turbulence driving strength, for $t > 30t_s$ the dissipation curves overlap. Additionally, the squares which mark $t = 100 \;\mathrm{Myr}$ are more or less at the same location across all four runs where the kinetic and thermal dissipation curves have overlapped. This would suggest $100$ Myr to be a "universal" turbulence dissipation timescale, depending on $n$ and $Z$ but fully indepedent from $\dot{\epsilon}_i$. This independence will also be explored further in Section \ref{sec:results:homo} on the dissipation of overdensities. The overlap also coincides with a plateau in thermal energy, formed by a combination of ineffficient cooling as the gas reaches $10^4$ K and the continuous cascade from kinetic to thermal energy.

We observe variations in the end state with respect to initial conditions in Figure \ref{enetimenonorm}, with all quantities unnormalized and curves separated by number density. The top thermal energy plot affirms the convergence of thermal energy curves observed in Figure \ref{enephasebig} across a larger, lower resolution sample size, with varied turbulence driving across fixed initial conditions. Since the metallicity is fixed, the spread within each $n$ bin represents only variations in $\dot{\epsilon}_i$. We denote the temperature floor with the black dotted line as an unphysical asymptote the thermal energy approaches across all runs, but as seen in Figure \ref{fig:cool}, the cooling rates drop by many orders of magnitude below $10^4$ K, so the actual physical evolution of thermal energy would not be a significant downwards deviation from a horizontal asymptote. We see that this convergence is independant of turbulence driving as the error regions decrease in size, and the value of $\epsilon_g$ at which the convergence begins, depends on $n$ and $Z$ (since inefficient cooling can also arise from low $n$ and/or low $Z$). We also affirm that these convergences occur at thermal energy plateaus after the initial rapid dissipation epoch. Kinetic energy, as with Figure \ref{enephasebig} and Figure \ref{enetime} also shows convergence and overlap. This convergence and overlap via binning both in $n$ and in $\dot{\epsilon}_i$, suggests a universal kinetic energy dissipation timescale. We note that this convergence point is at $100$ Myr in both Figures \ref{enetime} and \ref{enetimenonorm}, the same location marked by the squares in Figure \ref{enephasebig}.

The spatial scaling of the turbulence can be analyzed via
the turbulence power spectrum \citep{1982JGR....87.3617B,
  2015ApJ...808...48B, 2023ApJ...955...64B}, shown in Figure
\ref{powspecspline}. We compute the turbulence power spectra
as follows:
\begin{equation}
  \label{eqn:turbpowspec}
  P_k(k) = \dfrac{1}{2} k^2 \int d\Omega_k |\hat{v}(\vec{k})|^2
\end{equation}
where $\int d\Omega_k$ represents a solid angle integral
over $k$-space, and $k$ represents the radial component of
$\vec{k}$. $\hat{v}(\vec{k})$ represents the Fourier
Transform (FT) of the velocity field: 
\begin{equation}
  \label{eqn:velfourier}
  \hat{v}(\vec{k}) = \dfrac{1}{(2 \pi)^3} \int
  \vec{v}(\vec{r}) e^{-i\vec{k}\cdot\vec{r}} d^3r 
\end{equation}
The factor of $1/2$ in eq.~\eqref{eqn:turbpowspec}
represents the conversion from velocity-squared to specific
kinetic energy. 
 
The dissipation is visible in the downward shift in the
power spectra in the bottom plot, with relatively even
spacing between the power spectra at $t_s$, $10t_s$ and
$100t_s$. The upwards bump at higher $k$ along the inertial
regimes shows the bottleneck effect
\citep{PhysRevE.68.026304, 2010JFM...657..171D},
representing a buildup of energy below the dissipation
range. The inertial regimes of our runs match the kolmogorov
$k^{-5/3}$ power law \citep{1941DoSSR..30..301K}fairly well,
both during and after steady state turbulence. This is
indicative of the weakly supersonic nature of our runs with
$\mathcal{M}$ only being on order unity
\citep{2002A&A...390..307O,1981MNRAS.194..809L}. Notably,
stronger turbulence driving does not lead to an increasingly
shocked gas with a $k^{-2}$ power law
\citep{Burgers1948AMM}, but rather a near-congruent upwards
shift in the entire spectrum. The shape of the spectrum
remains preserved during dissipation, consistent with
\citep{2009A&A...504...33V}.

\subsection{Comparisons with Isothermal, Compressible GMC
  Dissipation}

\begin{figure}[h!]
  \centering
  \incgraph{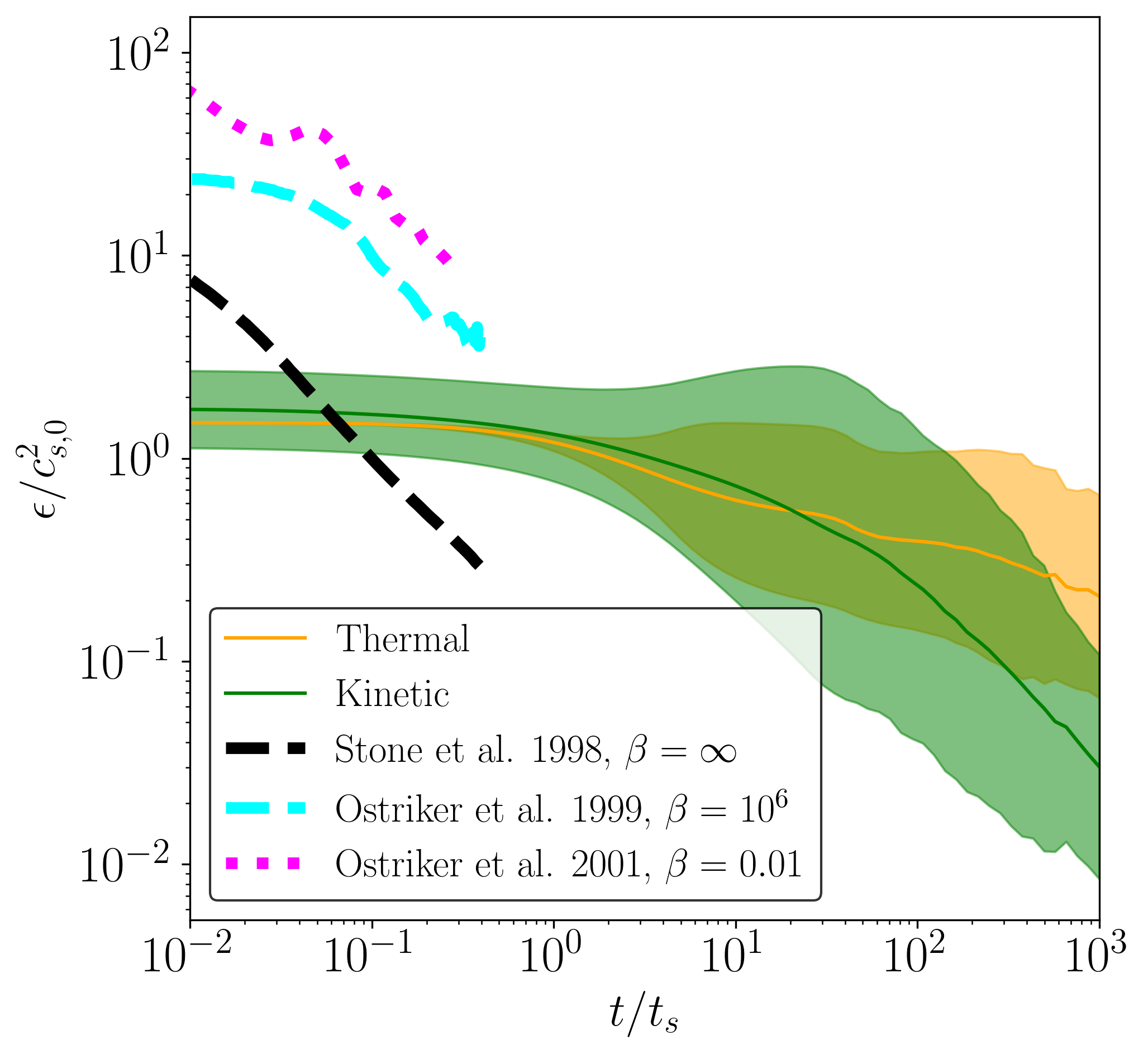}
  \caption{A comparison between our thermal and kinetic
    dissipation curves and the kinetic dissipation curves
    of \citet{1998ApJ...508L..99S} (black),
    \citet{1999ApJ...513..259O} (cyan) and
    \citet{2001ApJ...546..980O} (magenta). The latter
    three curves show dissipation under artificial
    viscosity, an isothermal equation of state and the
    presence of magnetic fields where
    $B = \beta^{-1/2} \times 1.4 \; (T/10 \;
    \mathrm{K})^{1/2} (n/10^2 \; \mathrm{cm}^{-3}) \; \mu
    \mathrm{G}$ (so $\beta = \infty$ represents pure
    hydrodynamics). The cyan and magenta dissipation
    curves include self-gravity. $t_s$ represents the
    kinetic energy saturation time defined as
    $\epsilon_k / \dot{\epsilon}_i$ during steady state
    turbulence, and the energies are normalized by sound
    speed squared $c_{s,0}^2$ right at turbulence driving
    turnoff.}
  \label{stonecompare}
\end{figure}
  
We compare our dissipation curves with various kinetic
dissipation curves in GMCs in Figure \ref{stonecompare}. The
black curve from \citet{1998ApJ...508L..99S} serves as a
fiducial comparison representing dissipation with pure
hydrodynamics without magnetic fields or self gravity. Two
key distinctions emerge - as also seen in Figure
\ref{eneloss_mach} our supersonic runs are only weakly
supersonic with $M$ on order unity, while
\citet{1998ApJ...508L..99S} sees stronger supersonic motions
with $M$ being an order of magnitude higher. The dissipation
timescales are also significantly shorter relative to $t_s$,
where kinetic energy dissipates well within $t_s$ while for
our runs both thermal and kinetic energy dissipate on
timescales greater than $10t_s$. Both distinctions are
reflective of the different astrophysical medium of interest
(GMCs are much cooler than CGMs, which allows significantly
weaker velocity perturbations to drive stronger supersonic
turbulent motions) and of the different physical processes
through which dissipation occurs (energy cascade and
radiative cooling for us, artificial viscosity for
\citet{1998ApJ...508L..99S}). The cyan and magenta curves
from \citet{1999ApJ...513..259O} and
\citet{2001ApJ...546..980O} illustrate the effects of self
gravity (both) and magnetic fields (latter). While they play
a minor role in stabilization against dissipation, their
dissipation timescales remain well within $t_s$.

We note that the artificial viscosity employed by the
aforementioned authors via the code \textsc{ZEUS}
\citep{1992ApJS...80..753S} is necessary to capture and
thermalize shocks, while \kratos adopts higher order Godunov
solver. Physically, in our work, kinetic dissipation occurs
only via shock thermalization through adiabatic
compression. Numerical viscosity also plays a role in shock
thermalization \citep{2013MNRAS.429.3353N}, though its
effects diminish with more accurate solvers and higher
resolution grids \citep{PhysRevE.68.046709}. Hearkening back
to Figure \ref{powspecspline}, it's likely that the
bottleneck \citep{PhysRevE.68.026304} is the result of both
a lack of subgrid viscous forces accounting for dissipation
at kolmogorov microscales, and insufficient numerical
viscosity given our numerical scheme and resolution.

\subsection{Energy Dissipation Timescales} \label{dissdata}
\begin{figure*}
    \centering
    \incgraph{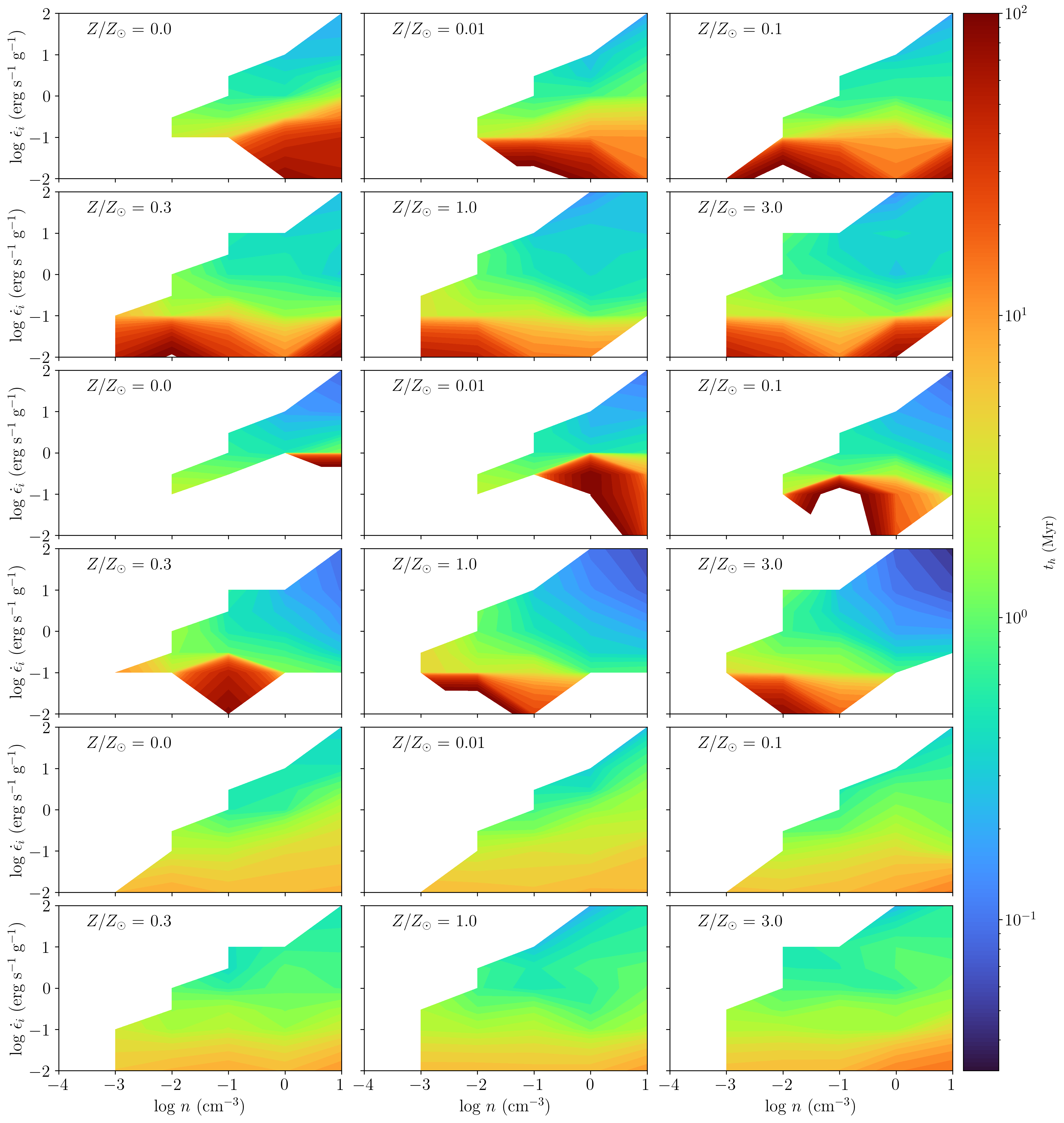}
    \caption{2D contour maps of the half energy dissipation timescales $t_h$, defined as $\epsilon(t_h) = 0.5\epsilon_0$, across only supersonic runs. The blank regions on the top left above the diagonal represent subsonic runs, and the blank regions on the bottom represent supersonic runs which did not dissipate half the component of energy in question. The first and second rows shows total energy, the middle third and fourth rows show thermal energy, and the bottom fifth and sixth rows show kinetic energy}
    \label{disstimescales}
\end{figure*}

We quantify the raw dissipation timescales $t_h$ in Figure \ref{disstimescales}, defined as the time it takes for the total, thermal or kinetic energy to drop by half. The primary trend on the top plot shows decreasing dissipation timescales with increasing $\dot{\epsilon}_i$ across all metallicities, with some $n$ dependence for higher metallicities particularly at $Z/Z_\odot = 1$ and $Z/Z_\odot = 3$. There is little discernible correlation between the dissipation timescale and metallicity -  rather, increased metallicity leads to enhanced cooling, allowing for some runs to reach the supersonic regime which would have been subsonic at lower metallicities, as seen with the increased number of supersonic runs above the diagonal for higher metallicity contour plots. On the middle and bottom plots, we can affirm our earlier observations from Section \ref{sec:results:thermalvskinetic}. Noting that blank bins below the diagonal in the thermal dissipation timescale represent supersonic runs that could not dissipate half their thermal energy over the simulation and interpreting those bins has having longer dissipation timescales, we see a steep drop in $t_h$ with increased $\dot{\epsilon}_i$ and $n$ of several orders of magnitude. Meanwhile in the bottom plot, while a $\dot{\epsilon}_i$ dependence in the kinetic dissipation timescales can be observed, it is far less steep and spans fewer orders of magnitude compared to the thermal dissipation timescales. No clear $n$ dependence can be observed either. The subsonic runs occupy the top left regions of all plots, above the diagonal, representing initial parameters where cooling is insufficient in preventing the gas from reaching a hot subsonic turbulent state. 
\begin{figure}
    \centering
    \incgraph{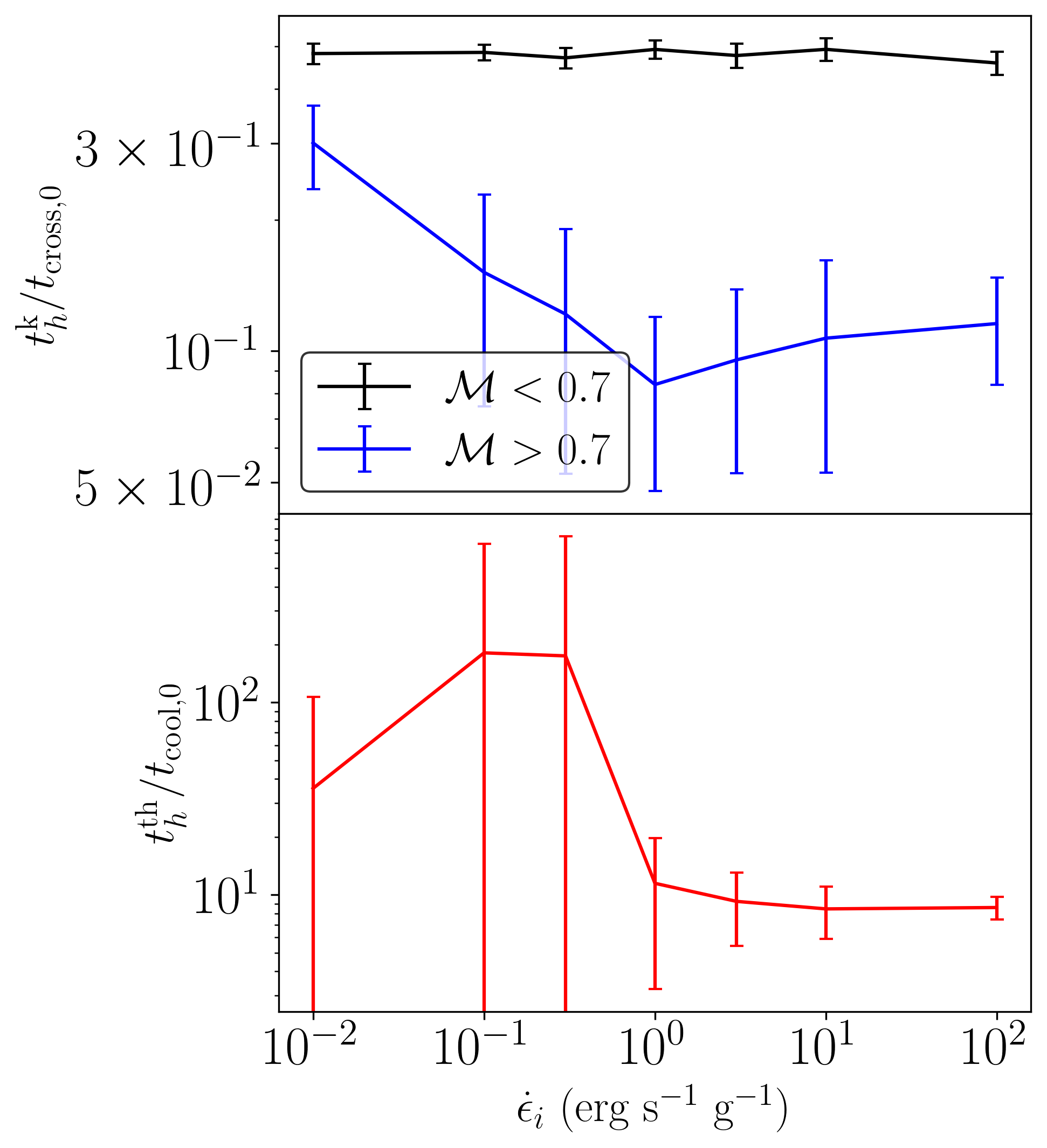}
    \caption{Dimensionless dissipation timescales $t_h$, defined as $\epsilon(t_h) = 0.5\epsilon_0$, for both kinetic (top) and thermal (bottom) energy as a function of $\dot{\epsilon}_i$. Kinetic energy is normalized by $t_\mathrm{cross,0}$ while thermal energy is normalized by $t_\mathrm{cool,0}$. Kinetic dissipation timescales are additionally separated between supersonic (blue) and subsonic (black) runs.}
    \label{disstableplot}
\end{figure}

We also present a set of dimensionless dissipation
timescales in Figure \ref{disstableplot}. We define a
kinetic timescale in
$t_\mathrm{cross,0} = \ell / \langle v_\mathrm{rms} \rangle$
and a thermal timescale in
$t_\mathrm{cool,0} = E_{g,0} / \sum^i n_i^2 \Lambda(T_i)
\Delta x_i^3$. The former represents the initial crossing
time at turbulence turnoff, where
$\langle v_\mathrm{rms} \rangle$ represents the mass
averaged rms velocity across every cell, while the latter
represents the initial cooling time at turbulence turnoff,
where $E_{g,0}$ represents the total initial thermal energy
and the denominator represents the total cooling rate
integrated across every cell.  Subsonic kinetic dissipation
shows a universal dimensionless timescale of approximately
$0.6 t_\mathrm{cross,0}$, while supersonic thermal
dissipation shows a convergence towards a universal
dimensionless timescale of roughly $8.5 t_\mathrm{cool,0}$.

\subsection{Density Homogenization}
\label{sec:results:homo}

\begin{figure}
  \centering
  \incgraph{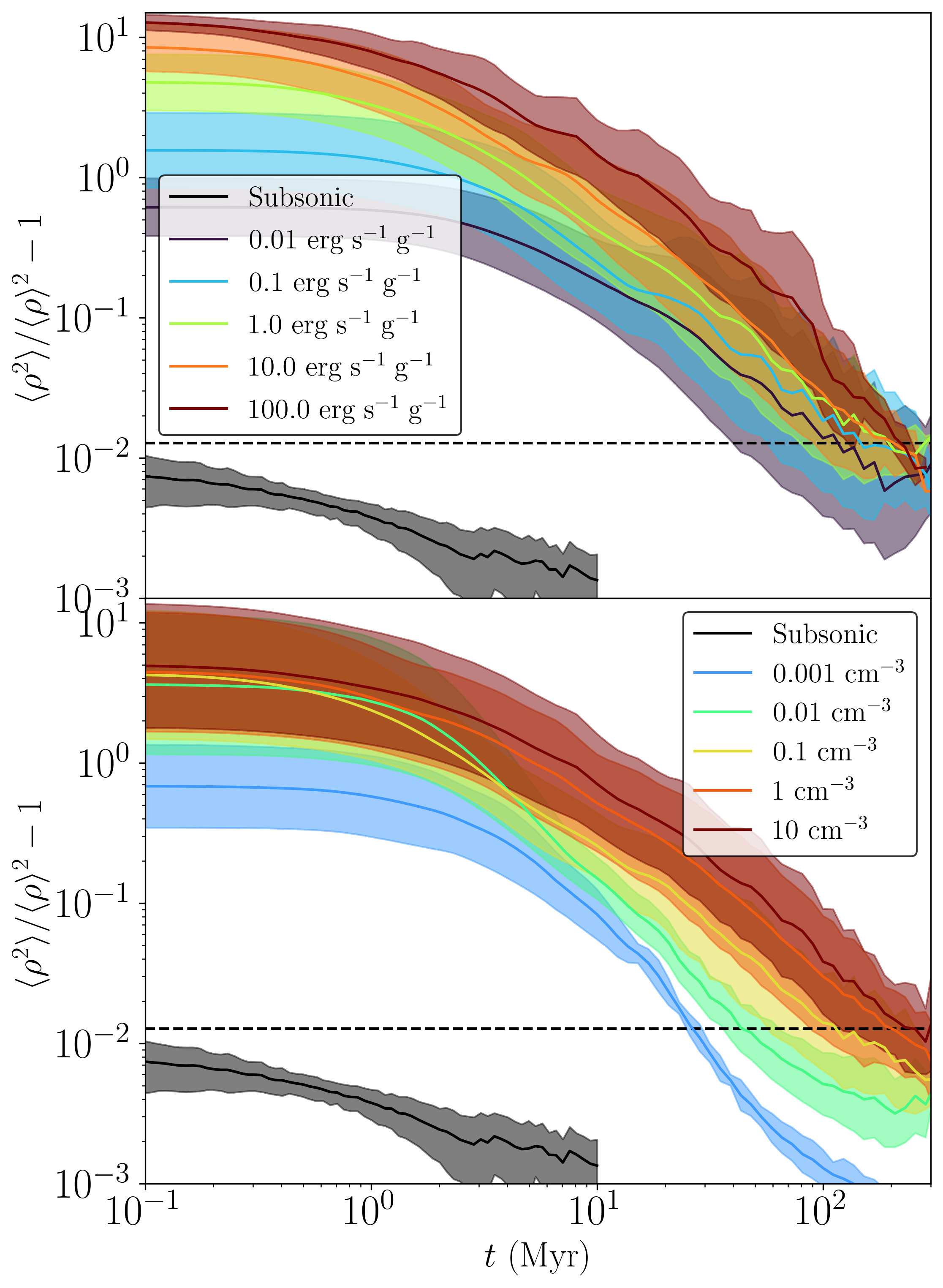}
  \caption{The dissipation of the clumping factor
    $\langle \rho^2 \rangle / \langle \rho \rangle^2 - 1$
    over time. Consistent with previous plots, the
    coloured lines show various binnings in initial
    simulation parameters across supersonic runs. The
    black line and shaded region shows the average and
    $1\sigma$ across all subsonic runs, and the dotted
    line shows $5\sigma$ across all timesteps for all
    subsonic runs. \textit{Top}: Clumping factor
    dissipation binned in
    $\dot{\epsilon}_i$. \textit{Bottom}: Clumping factor
    dissipation binned in $n$}.
  \label{clumpcompare}
\end{figure}

As seen in Figure \ref{phasecompare} and as discussed in
previous sections, there are density contrasts spanning
multiple orders of magnitude during steady state
turbulence. We define overall scale of such density
contrasts via the clumping factor, which is defined as
$C = \langle \rho^2 \rangle / \langle \rho \rangle^2$, where
$\langle \rho \rangle$ represents a spatial average across
all cells. We can examine the broad density homogenization
of the medium via dissipation in the clumping factor.  We
label the full definition of $C$ in subsequent plots and
figures for clarity, and plot $C - 1$ given $C = 1$
represents a completely uniform medium.

\begin{figure}
    \centering
    \incgraph{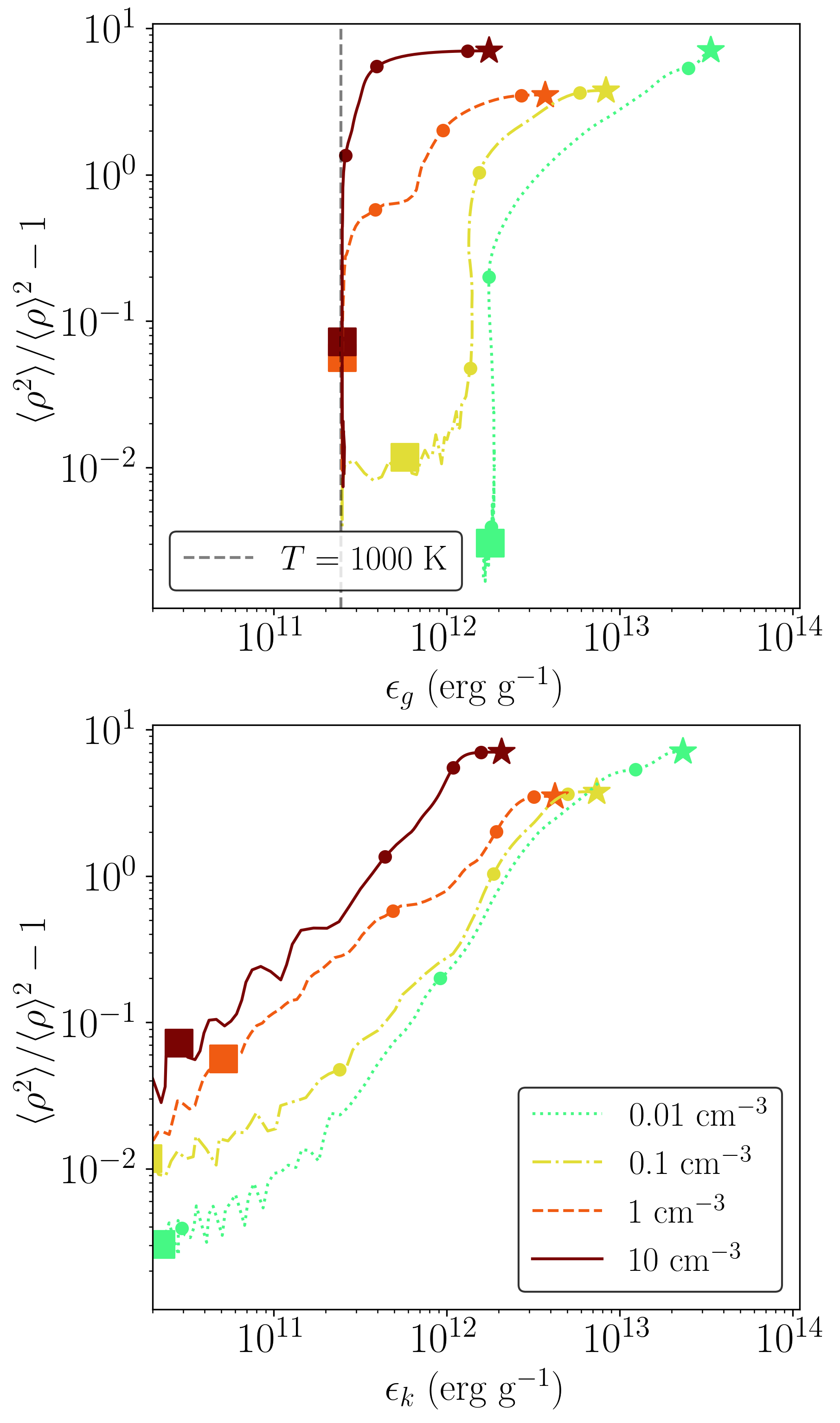}
    \caption{Phase plots tracking the temporal evolution of specific thermal (Top), specific kinetic energy (Bottom) relative to clumping factor. Four $128^3$ resolution runs are shown, corresponding to $Z/Z_\odot = 0.3$, $\dot{\epsilon}_i = 1 \; \mathrm{erg} \; \mathrm{s}^{-1} \; \mathrm{g}^{-1}$ and  $n$ ranging from $0.01 \; \mathrm{cm}^{-3}$ to $n = 10 \; \mathrm{cm}^{-3}$. Similar to Figure \ref{enephasebig}, stars denote the turbulence driving turn off time where each curve begins, circular dots denote $t_s$, $10t_s$ and $100t_s$ as one moves from the star and along the curve, and squares denote the 100 Myr mark. The black dotted line in the top plot marks the $T = 1000 \;\mathrm{K}$ point.}
    \label{clumpvsene}
\end{figure}
We show the time evolution of $C$ in Figure \ref{clumpcompare} with respect to both $\dot{\epsilon}_i$ and $n$.  An expected observation of the top plot shows a positive correlation between $\dot{\epsilon}_i$ and steady state clumping, as well as a significant valley between the clumping factors of subsonic and supersonic runs. Subsonic clumping factors do show a decrease of around a single dex, although this is insignificant compared to the several-dex decreases in the $C$ of the supersonic runs, especially when considering the $-1$ shift in the y axis. We define a supersonic tubulent medium to be homogenized if its clumping factor falls below a density homogenization limit of $C_\mathrm{sub} \approx 1.0132$ representing $5\sigma$ from the mean inital subsonic clumping factor. Functionally, this definition translates to a density homogenization timescale $t_\mathrm{Diss}^{\mathrm{C}}$, beyond which the density contrasts of a dissipated supersonic turbulent gas becomes indistinguishable from a subsonic turbulent gas. 

Despite the clear gradient in inital clumping factor positively correlated with $\dot{\epsilon}_i$, the curves and error regions converge and reach $C_\mathrm{sub}$ on roughly the same timescales of around 100 Myr, albeit with some spread between $30$ and $300$ Myr. When binned in $n$ on the bottom plot, a positive $n$ dependence in $t_\mathrm{Diss}^{\mathrm{C}}$ emerges. The shrinkage of the error regions encapsulates the same variations in $\dot{\epsilon}_i$ converging in the top plot. Hearkening back to Figures \ref{enetime} and \ref{enetimenonorm}, dissipation in $C$ is similar to but not congruent to dissipation in $\epsilon_k$. While both quantities dissipates exponentially, kinetic energy dissipation does not depend on $n$.  

We examine the intercorrelation between clumping factor dissipation and energy dissipation more closely in Figure \ref{clumpvsene}. Since the locations of the squares denote a fixed time $t = 100$ Myr, the lower the position of the square, the shorter $t_\mathrm{Diss}^{\mathrm{C}}$ is. We note that only the two lowest values of $n$ are remotely representative of conditions in the CGM, and our purpose of showing $n = 1 \; \mathrm{cm}^{-3}$ and $n = 10 \; \mathrm{cm}^{-3}$ is to illustrate a trend. Given their similar dissipative behaviours in Figures \ref{enetime} and \ref{clumpcompare}, the middle plot unsurpsingly shows a steady power law relation between $C$ and $\epsilon_k$. The correlation with thermal dissipation in the top plot is much less smooth, and mirrors the the time tracks in Figure \ref{enephasebig}. $C$ only shows a power law relation with $\epsilon_g$ during the rapid thermal dissipation epoch, continues to dissipate during the $\epsilon_g$ plateau. There is a nonlinear relation between initial $C$ and $n$, with $n = 0.01 \; \mathrm{cm}^{-3}$ and $n = 10 \; \mathrm{cm}^{-3}$ exhibiting higher clumping factors than the other runs. We again observe and affirm the negative correlation in the top plot between $n$ and $C$ at 100 Myr as seen in Figure \ref{clumpcompare}, but we also note the dissipated steady state behaviour of $\epsilon_g$. The more diffuse runs with lower clumping factors at 100 Myr have hotter thermal energy plateaus, with the $n = 0.01 \; \mathrm{cm}^{-3}$ and $n = 0.1 \; \mathrm{cm}^{-3}$ not even reaching the 1000 K temperature floor. 

\begin{figure}
    \centering
    \incgraph{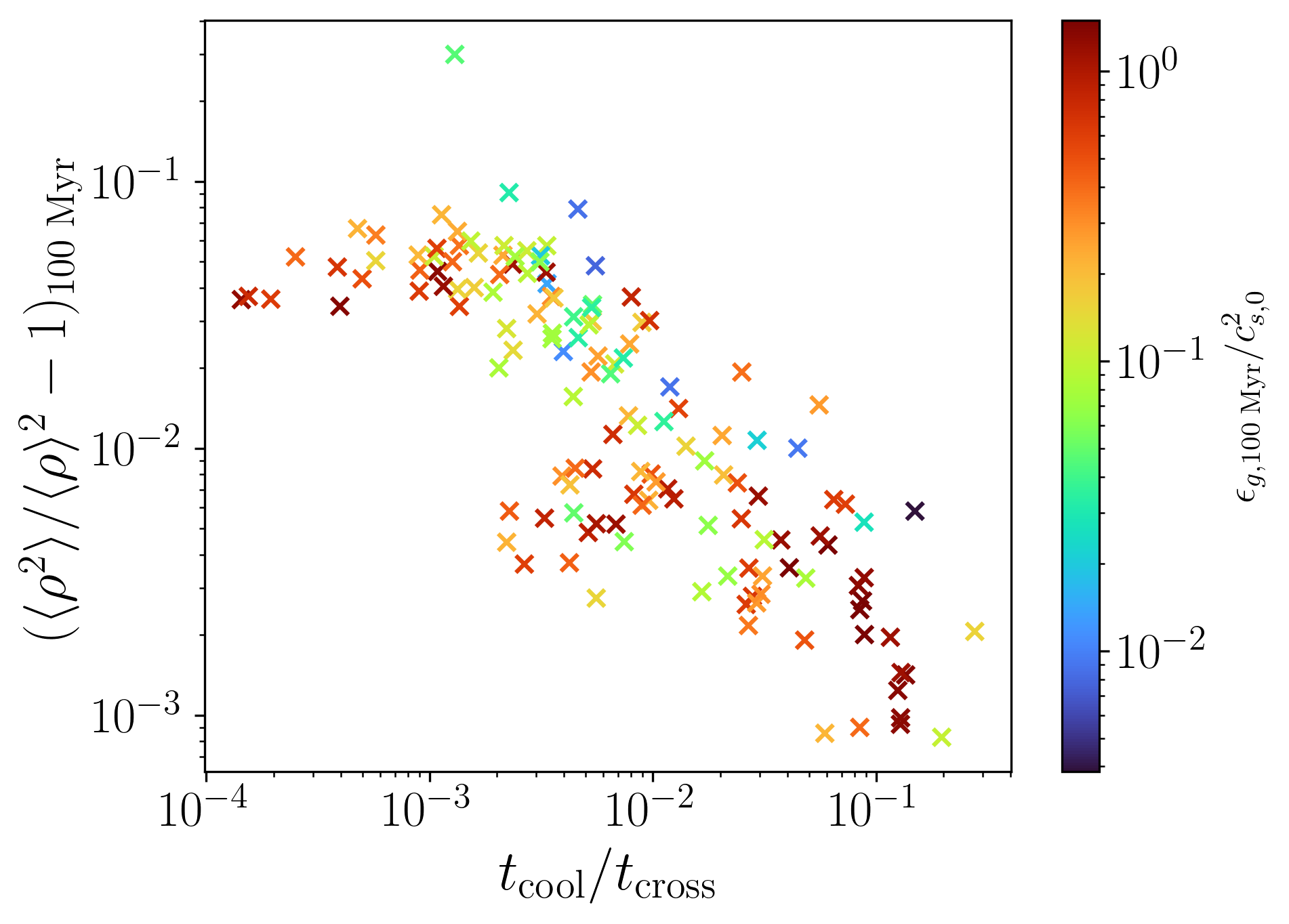}
    \caption{The ratio between $t_\mathrm{cool}$ and $t_\mathrm{cross}$ at turbulence turnoff versus the clumping factor $\langle \rho^2 \rangle / \langle \rho \rangle^2 - 1$ at 100 Myr. Each point represents a supersonic turbulent run, with the colour bar and marker colours denoting the normalized thermal energy $\epsilon_g/c_{s,0}^2$ at 100 Myr. As with previous figures, $c_{s,0}$ represents the sound speed at turbulence turnoff.}
    \label{tcoolvsclump}
\end{figure}

We affirm this trend holds statistically via an examination across all supersonic turbulent runs in Figure \ref{tcoolvsclump}, and it becomes evident the $n$ dependence of $t_\mathrm{Diss}^C$ represents a more fundamental positive dependance on $t_\mathrm{cool}/t_\mathrm{cross}$. The more inefficient the cooling, the more uniform the medium becomes. There are no discernable trends in the colouring of the data points, showing that neither $t_\mathrm{cool}/t_\mathrm{cross}$ nor $t_\mathrm{Diss}^C$ correlate significantly with the process of thermal dissipation. Rather, the dependence of $t_\mathrm{Diss}^C$ on $t_\mathrm{cool}/t_\mathrm{cross}$ reflects a dependence on the current dynamical state of the gas. Going back to Figure \ref{clumpvsene}, despite having similar kinetic energies at 100 Myr, the $n = 0.01 \;\mathrm{cm}^{-3}$ run has distinctly more thermal energy than the $n = 10 \; \mathrm{cm}^{-3}$ run at that time, representing a smaller $t_\mathrm{cool}/t_\mathrm{cross}$ and hence a shorter density homogenization timescale, as the gas cannot cool fast enough to prevent the bulk motions from diffusing and smoothing out overdensities.

\subsection{Fourier Analysis of Turbulent Clouds} \label{sec:cloudcrush}
In this section we characterize the turbulence driven overdensities and clouds in more detail. While $C$ can broadly describe the overall "clumpiness" of the medium, it is insufficient in characterizing properties such as the spatial scales or $\mathcal{M}$ of the substructures. We extend our power spectrum analysis from Figure \ref{powspecspline} to more gas properties, following similar methods to those of \cite{2013ApJ...763...51F}. For some field $q(\vec{r})$ over $\mathbb{R}^3$, its power spectrum is defined as
\begin{equation}\label{eqn:generalpowspec}
P_q(k) = k^2 \int d\Omega_k |\hat{q}(\vec{k})|^2
\end{equation}
where $\int d\Omega_k$ represents a solid angle integral over $k$-space, and $k$ represents the radial component of $\vec{k}$. $\hat{q}(\vec{k})$ represents the Fourier Transform (FT) of the field $q$:
\begin{equation} \label{eqn:generalfourier}
\hat{q}(\vec{k}) = \dfrac{1}{(2 \pi)^3} \int q(\vec{r}) e^{-i\vec{k}\cdot\vec{r}} d^3r
\end{equation}

\begin{figure*}
    \centering
    \incgraph{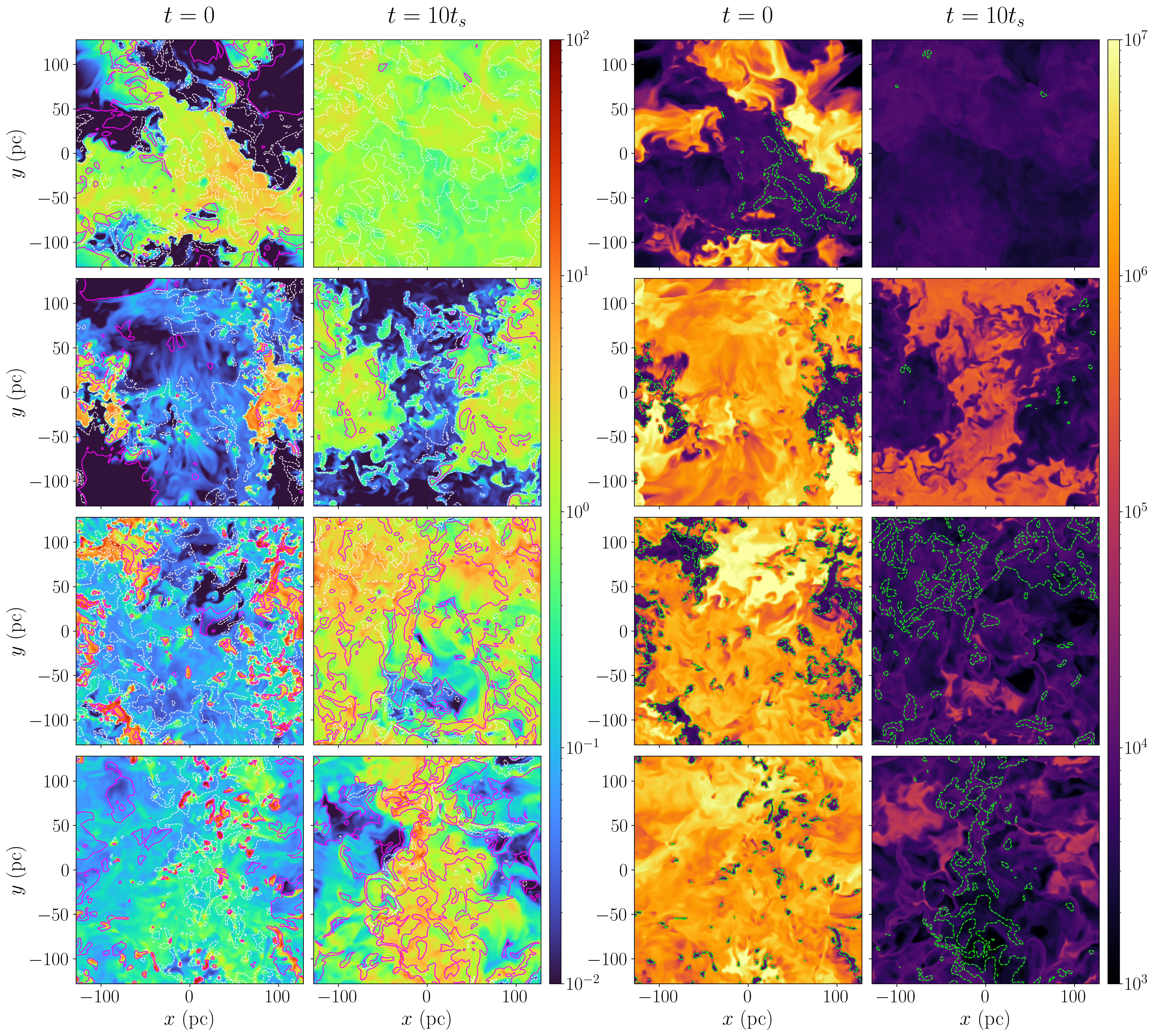}
    \caption{Density (Left) and temperature (Right) slice plots of four $256^3$ resolution runs corresponding to $Z/Z_\odot = 0.3$, $n = 0.1 \; \mathrm{cm}^{-3}$. Each row corresponds to a different value of $\dot{\epsilon}_i$ - from top to bottom they are: $0.1 \; \mathrm{erg} \; \mathrm{s}^{-1} \; \mathrm{g}^{-1}$,  $1 \; \mathrm{erg} \; \mathrm{s}^{-1} \; \mathrm{g}^{-1}$,  $3 \; \mathrm{erg} \; \mathrm{s}^{-1} \; \mathrm{g}^{-1}$ and $10 \; \mathrm{erg} \; \mathrm{s}^{-1} \; \mathrm{g}^{-1}$. The densities and colour bar are normalized by the initial density $n = 0.1 \; \mathrm{cm}^{-3}$. The columns show the gas at different times, the first and third columns right at turbulence turnoff, the second and fourth columns at $10t_s$. On the density slice plots, dashed white lines show $\mathcal{M} = 1$ contours and solid magenta lines show $\mathcal{M} = 3$ contours. On the temperature slice plots, dashed lime lines show $\rho/m_p/n = 3$ contours .}
    \label{densmacharray}
\end{figure*}

Figure \ref{densmacharray} paints a physical picture of the gas during steady state turbulence and during dissipation. While it's evident from Figure \ref{clumpcompare} there should be higher density contrasts with higher $\dot{\epsilon}_i$, another effect of stronger driving is a dramatic shrinking in the typical clump sizes. The $\dot{\epsilon}_i = 1 \; \mathrm{erg} \; \mathrm{s}^{-1} \; \mathrm{g}^{-1}$ run sees clumps of order $10^2$ pc, while the $\dot{\epsilon}_i = 10 \; \mathrm{erg} \; \mathrm{s}^{-1} \; \mathrm{g}^{-1}$ run sees clumps of order $10^0$ to $10^1$ pc in size. There is a nonlinear though clear inverse relation between driving strength and clump size. Weak turbulence driving, as seen in the $\dot{\epsilon}_i = 0.1 \; \mathrm{erg} \; \mathrm{s}^{-1} \; \mathrm{g}^{-1}$ run, sees diffuse "bubbles" rather than dense clumps. The $\mathcal{M} = 1$ contours weakly trace the boundaries between diffuse and dense regions, and whose ubiquity shows both regions to be broadly supersonic. The $\mathcal{M} = 3$ contours show even more limited overlap with overdense regions in the top two plots, but trace dense regions very well in the bottom two plots. Only under strong turbulence driving where $\dot{\epsilon}_i \geq 3 \; \mathrm{erg} \; \mathrm{s}^{-1} \; \mathrm{g}^{-1}$ do we observe strongly supersonic clumps. In the temperature plots on the two left columns, the cool regions correlate very well with the dense regions on the two right columns, and share similar temperatures of $10^3$ to $10^4$ K across all four runs. This would suggest that faster bulk motions within the clumps is a significant contributor towards their increased compressibility.

\begin{figure*}
    \centering
    \incgraph{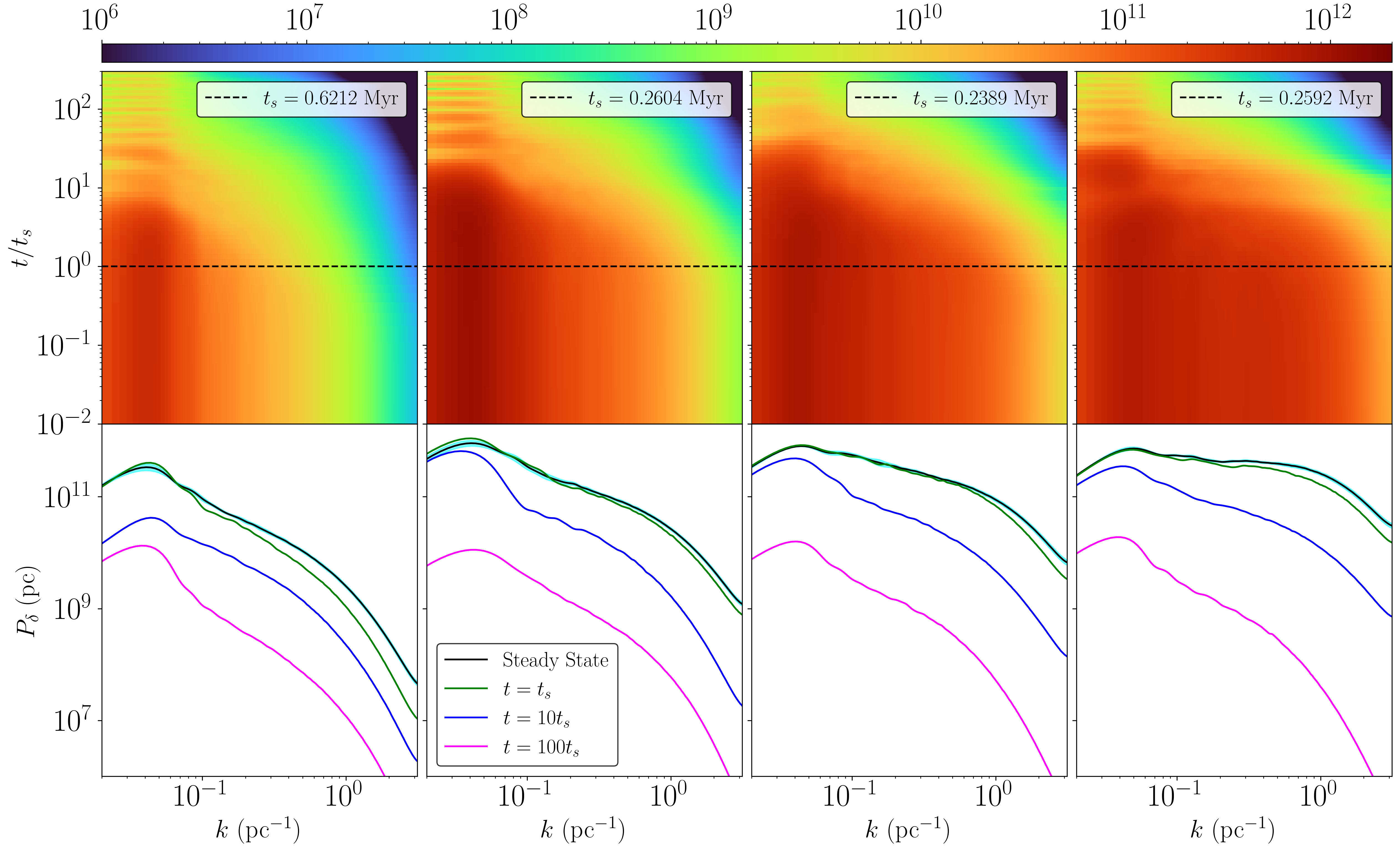}
    \incgraph{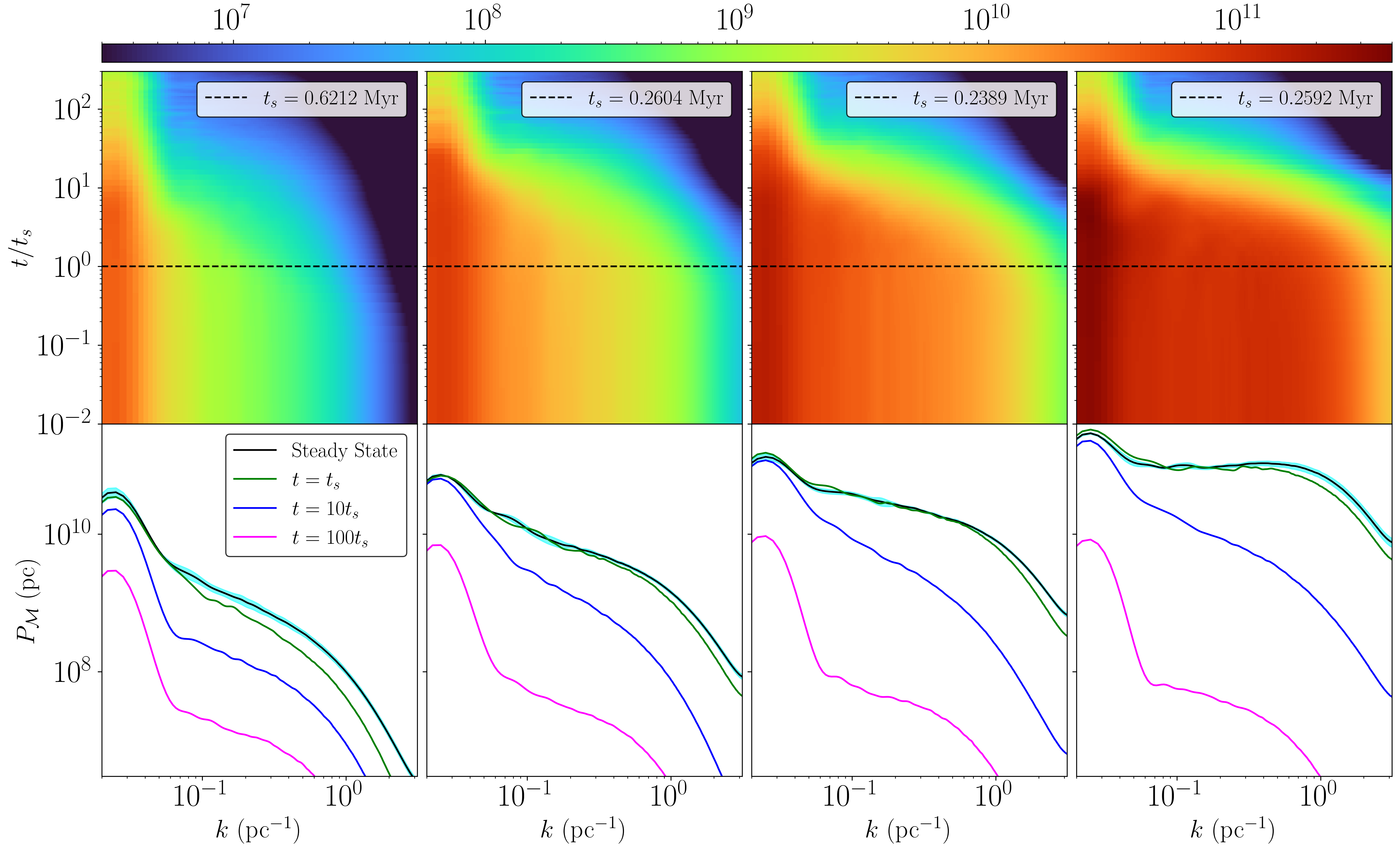}
    \caption{The density (Top) and $\mathcal{M}$ (Bottom) power spectra of the four high resolution $256^3$ runs shown in Figure \ref{enephasebig}, where $Z/Z_\odot = 0.3$ and $n = 0.1 \; \mathrm{cm}^{-3}$, and $\dot{\epsilon}_i$ being 0.1, 1, 3.0 and 10.0 $\mathrm{erg} \; \mathrm{s}^{-1} \; \mathrm{g}^{-1}$ from left to right. The bottom row plots the turbulence power spectra during steady state turbulence, after $t_s$, $10t_s$ and $100t_s$. The cyan coloured region represents $1\sigma$ across the last twenty outputs of the steady state turbulence epoch. The top row plots the power spectra from turbulence turn off up to $100t_s$, with the colour denoting $\epsilon_k$. The value of $t_s$ is also shown across each of the four runs. Both the top and bottom row plots are smoothed using degree-5 polynomial splines for visualization purposes. }
    \label{mattermachpowspecspline}
\end{figure*}

We charactize the spatial scalings of $\mathcal{M}$ via a power spectrum of the mach number. Following equations~\eqref{eqn:generalpowspec} and \eqref{eqn:generalfourier} we define $P_q(k) = P_\mathcal{M} (k)$ and $\hat{q}(\vec{k}) = \hat{\mathcal{M}}(\vec{k})$ from $q(\vec{r}) = \mathcal{M}(\vec{r})$, where $\mathcal{M}(\vec{r})$ represents the spatial mach number define as:
\begin{equation} \label{eqn:mach}
	\mathcal{M}(\vec{r}) = \dfrac{|\vec{v}(\vec{r})|}{c_s(\vec{r})}
\end{equation}
Similarly we an characterize the spatial scalings of the overdensities using the matter power spectrum $P_\delta(k)$, for which $q(\vec{r}) = \delta(\vec{r})$ where $\delta$ is a dimensionless overdensity parameter defined as 
\begin{equation}
\delta(\vec{r}) = \dfrac{\rho(\vec{r}) - \bar{\rho}}{\bar{\rho}}
\end{equation}
where $\bar{\rho}$ represents the spatially averaged density across the box, although functionally this equates to the initial density of the run.

In Figure \ref{mattermachpowspecspline}, we characterize the dissipation of the dense clumps as seen in Figure \ref{densmacharray} via the time evolution $P_\mathcal{M}$ and $P_\delta$ across the same four runs. Both power spectra exhibit broken $k^{-\alpha}$ power laws, with higher $\dot{\epsilon}_i$ corresponding to shallower $\alpha$ in the $0.07 < k < k_d$ range, where $k_d \approx 1 \; \mathrm{pc}^{-1}$ during turbulence driving. We will refer to this range in $k$ as the linear range, the slope of the linear range as $\alpha_\delta$ and $\alpha_\mathcal{M}$ for $P_\delta$ and $P_\mathcal{M}$ repectively, and $k > k_d$ as the dissipation range. The maxima of $P_\delta$ for $\dot{\epsilon}_i = 3 \; \mathrm{erg} \; \mathrm{s}^{-1} \; \mathrm{g}^{-1}$ and $\dot{\epsilon}_i = 10 \; \mathrm{erg} \; \mathrm{s}^{-1} \; \mathrm{g}^{-1}$ are around the same as if not lower than the maxima for $\dot{\epsilon}_i = 1 \; \mathrm{erg} \; \mathrm{s}^{-1} \; \mathrm{g}^{-1}$, representing a larger proportion of the overdensities being captured at smaller scales. A shallow $\alpha_\delta$ represents not just the presence of distinct clumps at small scales, but hierarchically structured clumping, where dense clumps host even denser clouds within. For $P_\mathcal{M}$, a shallower $\alpha_\mathcal{M}$ represents those same clumps being increasingly supersonic due to underpressurization from enhanced cooling. The rise in the linear range is also noticeably sharper from $\dot{\epsilon}_i = 3 \; \mathrm{erg} \; \mathrm{s}^{-1} \; \mathrm{g}^{-1}$ to $\dot{\epsilon}_i = 10 \; \mathrm{erg} \; \mathrm{s}^{-1} \; \mathrm{g}^{-1}$, which is consistent with Figure \ref{densmacharray}. 

During dissipation, $P_\delta$ sees a steady increase in $\alpha_\delta$ and decrease in $k_d$, which represents hierarchical dissipation of substructures beginning at the smallest scales. We see this in Figure \ref{densmacharray} (primarily for $\dot{\epsilon}_i \geq 1 \;\mathrm{erg} \; \mathrm{s}^{-1} \; \mathrm{g}^{-1}$), where at $10t_s$, the gas is smoother on smaller scales compared to how it is at $t_s$, while preserving its clumpiness on larger scales.  $P_\mathcal{M}$ dissipates more irregularly, given its strong coupling with cooling rate $\Lambda$. While $k_d$ also decreases, $\alpha_\mathcal{M}$ evolves very differently when compared with $\alpha_\delta$. For $\dot{\epsilon}_i = 0.1 \; \mathrm{erg} \; \mathrm{s}^{-1} \; \mathrm{g}^{-1}$, there is no significant change in $\alpha_\mathcal{M}$, while for the three runs with higher $\dot{\epsilon}_i$, $\alpha_\mathcal{M}$ increases between $t = 0$ and $t = 10t_s$ and decreases between $t = 10t_s$ and $t = 100t_s$. The drop in amplitude is more significant than that of $P_\delta$ and scales with increasing $k$, where cooling becomes increasingly dominant. 

\begin{figure}
    \centering
    \incgraph{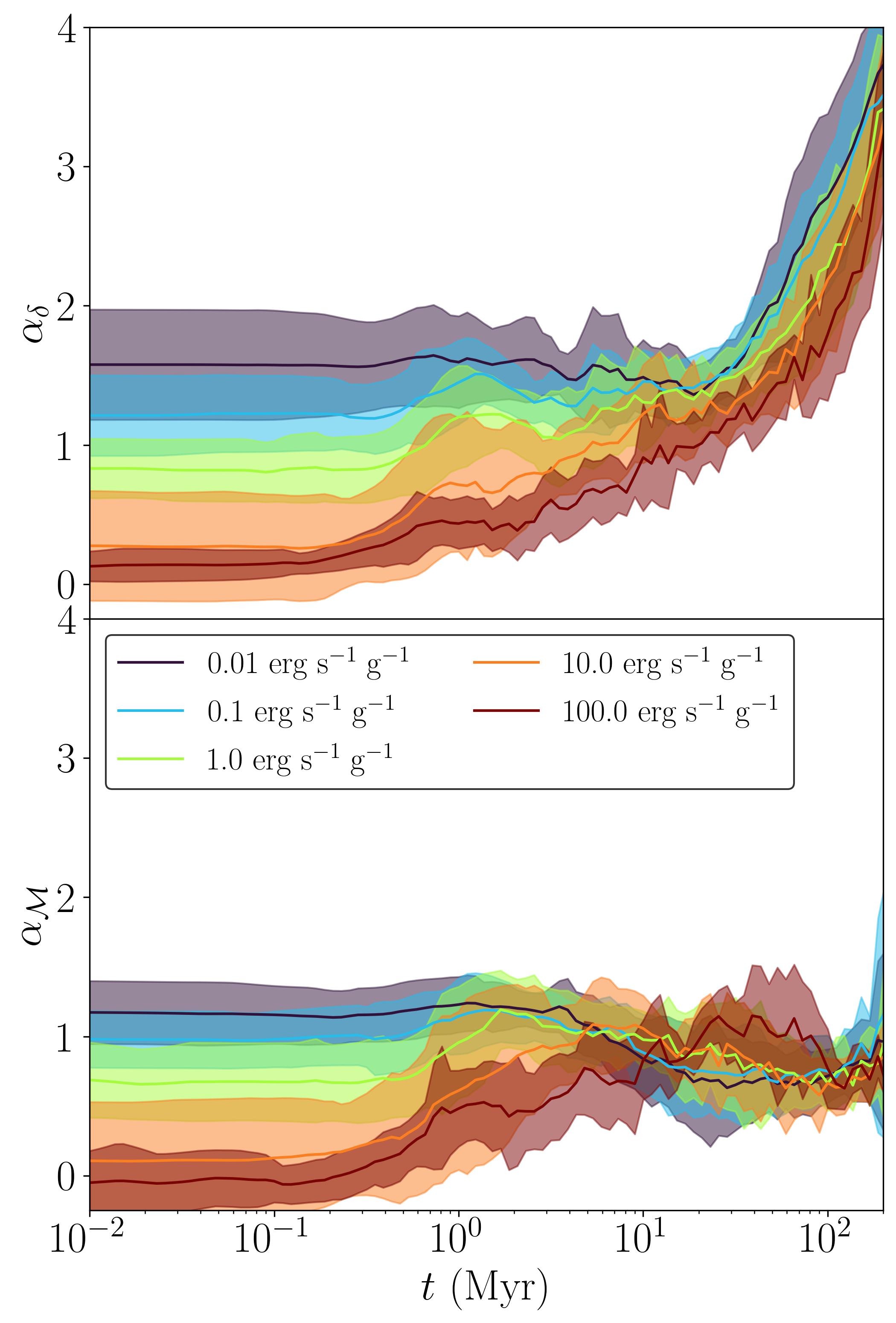}
    \caption{The power law indices of $P_\mathcal{M}$ (Top) and $P_\delta$ (Bottom) over time across supersonic $128^3$ runs. The linear range power laws are separated in $\dot{\epsilon}_i$, and the dissipation range power laws are coagulated and shown in black. The shaded regions represent $1\sigma$.  }
    \label{powlawvtime}
\end{figure}

The shortened cloud survival timescales at efficient cooling regimes does not contradict existing evidence that radiative cooling supports cloud \citep{2018MNRAS.480L.111G,2020MNRAS.492.1970G,2020MNRAS.492.1841L}. Rather, it speaks on the typical sizes of such clouds. The dissipation of clouds due to shockwaves is governed by cloud-crushing timescale \citep{1994ApJ...420..213K} defined as 
\begin{equation} \label{eqn:cloudcrush}
t_\mathrm{cc} = \dfrac{\chi^{1/2} a_0}{v_b}
\end{equation}
where $\chi$ is the dimensionless cloud overdensity factor, $a_0$ the cloud radius and $v_b$ the incoming shock velocity. Clouds can only survive shocks when $t_\mathrm{cool} < t_\mathrm{cc}$ \citep{2009ApJ...703..330C,2018MNRAS.480L.111G,2020MNRAS.492.1970G}, when a significant pressure gradient can be maintained between the cloud and the intracloud medium. The destruction of clouds as seen in the steepening of $\alpha_\delta$ and the difference between the $t = 0$ and $t = 10t_s$ snapshots in Figure \ref{densmacharray} show that $t_\mathrm{cool} > t_\mathrm{cc}$ on all scales. Assuming a uniform "intracloud medium" shock speed $v_b$ when integrated across all clouds in all directions given our uniform randomized turbulence driving scheme (See Section \ref{sec:methods}), the typical cloud size must then have a sharp inverse dependence on $\chi$, where $a_0 \propto \chi^{-\beta}$ for $\beta > 3/2$ since $t_\mathrm{cool} \propto \chi^{-2}$. We note this as a generalized empirical result given the nonideal conditions of our simulations and our broader focus on global energy and substructure dissipation, whereas detailed studies on cloud survival and crushing employ idealized wind tunnel simulations  \citep{1994ApJ...420..213K,1995ApJ...454..172X,2009ApJ...703..330C,2018MNRAS.480L.111G,2020MNRAS.492.1970G,2020MNRAS.492.1841L,2021MNRAS.501.1143K}.

We compute $\alpha_\mathcal{M}$ and $\alpha_\delta$ via a curve fit using the Levenberg-Marquardt least-squares algorithm \citep{doi:10.1137/0111030} for $0.07 < k < k_d$. To account for the shrinking of $k_d$ over time, multiple curve fits are performed for each run at each timestep for $0.2 < k_d \pi/4$, and the fit with the smallest error is chosen. The bounds are chosen from visual inspection of the $128^3$ resolution power spectra, encapsulating the full range of $k_d$ across all runs and all times. 

\begin{figure*}
    \centering
    \incgraph{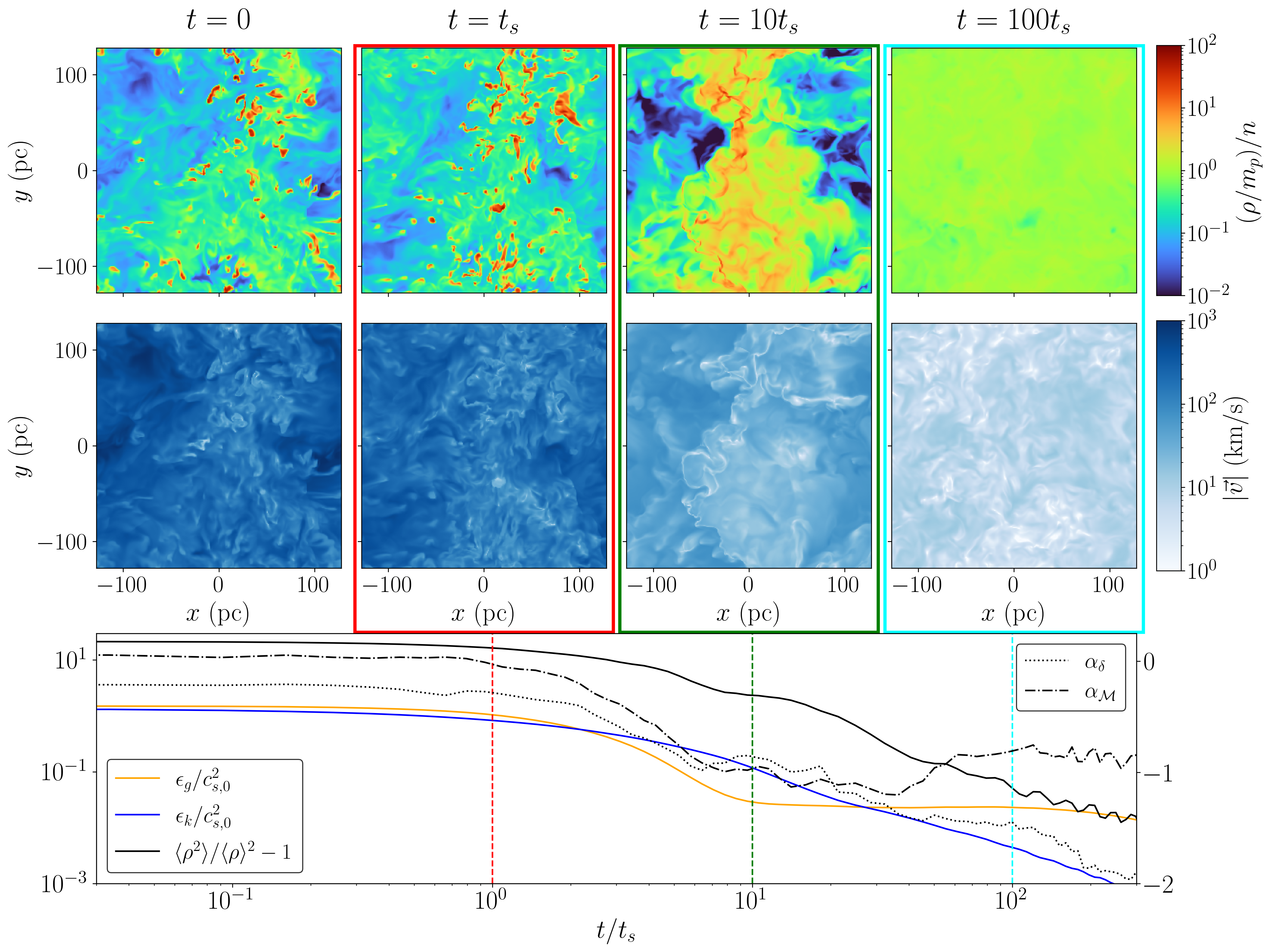}
    \caption{The full set of dissipation curves for the $Z/Z_\odot = 0.3$, $n = 0.1 \; \mathrm{cm}^{-3}$, and $\dot{\epsilon}_i = 10 \; \mathrm{erg} \; \mathrm{s}^{-1} \; \mathrm{g}^{-1}$. Across four distinct snapshots $t = 0, t_s, 10t_s, 100t_s$, the top row plots density slices normalized by $n$ while the middle row plots velocity magnitude. The bottom plot shows five dimensionless dissipation curves over time normalized by $t_s$: thermal energy ($\epsilon_g/c_{s,0}^2$), kinetic energy ($\epsilon_k/c_{s,0}^2$), clumping factor ($C = \langle \rho^2 \rangle / \langle \rho \rangle^2 - 1$), matter power spectrum linear regime power index ($\alpha_\delta$) and mach power spectrum linear regime power index ($\alpha_\mathcal{M}$). The former three quantities are shown on the left $y$ axis, while the latter two are shown on the right $y$ axis. The red, green and cyan dotted lines on the bottom plot corresponding to the boxes of the same colour around the slice plots show the times slice plots represent.}
    \label{totaltimeevolution}
\end{figure*}

We show the power indicies for $P_\mathcal{M}$ and
$P_\delta$ over time across all our supersonic $128^3$ runs
in Figure \ref{powlawvtime}. Both power spectra show an
initial gradient in $\alpha$, with higher $\dot{\epsilon}_i$
leading to shallower $\alpha$. The power law indices for
$P_\delta$ in the top plot initially converge up until $10$
Myr, beyond which they begin to sharply steepen. Meanwhile
for $P_\mathcal{M}$, the power law indicies also initially
converge up until the same time, but remain more or less
unchanged as a whole beyond that time, with
$\dot{\epsilon}_i = 100 \; \mathrm{erg} \; \mathrm{s}^{-1}
\; \mathrm{g}^{-1}$ serving as a mild outlier. Both
behaviours are consistent with Figure
\ref{mattermachpowspecspline}. As with $C$ in Figure
\ref{clumpcompare}, $\alpha$ shows distinct and synchronous
trends with respect to absolute time $t$, but unlike Figure
\ref{clumpcompare}, this no distinct $n$ dependence in the
time evolution of $\alpha$. The steepening of $\alpha$ in
the top plot represents the crushing of small scale clouds,
and from the convergent behaviour of the curves we see that
similar to $t^C_\mathrm{Diss}$, the cloud crushing
timescales are weakly coupled to if not independent from
$\dot{\epsilon}_i$. Similar to Figure \ref{clumpcompare}, we
can roughly define a universal cloud crushing timescale
$t_{cc}$ to be apprxoimately 10 Myr, representing the
convergence point in the power law indices across all
supersonic runs.

In Figure \ref{totaltimeevolution} we compare the time
evolution of $\alpha$ in relation to the other quantities
examined in this study, namely thermal energy, kinetic
energy, and clumping factor. The initial state is broadly
weakly supersonic where $\epsilon_g / \epsilon_k \approx 1$
with the clumps being strongly supersonic as seen from
Figure \ref{densmacharray}. $C$ and $\epsilon_k$ are more or
less parallel, affirming our earlier observations and
discussions. The bulk of the dissipation occurs between
$t_s$ and $10t_s$, where a sharp drop in $\epsilon_g$
corresponds to a steepening in $\alpha_\mathcal{M}$ and
$\alpha_\delta$. For a brief period of time between $3t_s$
and $20t_s$, the gas is strongly supersonic due to rapid
thermal cooling, and within this period we observe $C$,
$\alpha_\mathcal{M}$ and $\alpha_\delta$ to roughly flatten
out at various points. This may be indicative of temporary
stabilization against cloud collapse, where the clouds
become significantly larger and colder resulting in longer
$t_\mathrm{cc}$ and stronger pressure gradients between the
cloud and intracloud medium.

\section{Discussion}

\subsection{On the Nature of Supersonic Thermal and Kinetic
  Dissipation}

Thermal and kinetic energy exhibit vastly different
dissipative behaviours. The bulk of the thermal energy
dissipates during the initial rapid dissipation epoch while
kinetic energy maintains a consistent exponentially decaying
rate. The physics behind the rapid thermal dissipation epoch
results from enhanced cooling in dense clouds, evidenced by
stronger turbulence driving leading to a clumpier medium,
shallower $P_\delta$ power law and increased thermal
dissipation as seen in Figures \ref{clumpcompare},
\ref{mattermachpowspecspline} and \ref{enetime}. Thermal
dissipation is then strongly coupled to turbulence driving
indirectly through increased clumping. Kinetic dissipation
on the other hand occurs via energy cascade into thermal
energy, per the following relation in the inertial regime as
derived by \citet{1941DoSSR..30..301K}:
\begin{equation}
  \label{enecasc}
  E_k = C_k k^{-5/3} \dot{\epsilon}_k^{2/3}
\end{equation}
As seen in Figure \ref{powspecspline}, our simulations
follow the $k^{-5/3}$ power law quite well, despite the
bottleneck effect \citep{PhysRevE.68.026304}. We note that
since we cannot resolve the kolmogorov microscales and we
did not introduce additional subgrid artificial viscosity,
the cascade occurs via only shock heating and numerical
dissipation, which may have represented an underestimation
of up to 50\% \citep{1998ApJ...508L..99S}. In comparatively
stronger supersonic regimes, the power law approaches
$k^{-2}$ for a fully shocked gas \citep{Burgers1948AMM} and
the bottleneck effect becomes less significant
\citep{2010A&A...512A..81F,2013ApJ...763...51F}. While we do
achieve strongly supersonic regions within dense clumps,
overall our supersonic runs are weakly supersonic where
$\mathcal{M}$ and $\epsilon_k/\epsilon_g$ are of order unity
during steady state turbulence, due to shock heating of the
gas and inefficient cooling regimes beyond the local maximum
along $\Lambda$ beyond $10^4$ to $10^5$ K depending on
$Z/Z_\odot$.  s This kinetic dissipation rate
$\dot{\epsilon}_k^{2/3}$ depends only on $E_k$, which leads
to an indirect coupling with $\dot{\epsilon}_i$ where strong
turbulence driving leads to higher initial kinetic
energy. However this correlation only results in exponential
dissipation, which leads to convergence in the kinetic
energy curves across all turbulence driving strengths. We
observe similarities with \citet{1998ApJ...508L..99S} and
\citet{1999ApJ...513..259O} with the shapes of our kinetic
dissipation curves, although in our case it takes up to
$100t_s$ for kinetic energy to dissipate by 1 dex, while in
their case 1 dex dissipation occurs within $0.1
t_s$. Substructures, as the product of shock compression
during turbulence driving, naturally couple with kinetic
energy and dissipate similarly. Their lifetimes and
dissipation rate are also coupled with thermal dissipation,
since strong cooling and further underpressurization within
the cloud leads to increased cloud stability against
crushing \citep{2009ApJ...703..330C, 2018MNRAS.480L.111G,
  2020MNRAS.492.1970G}.

\subsection{On the Formation and Crushing of Turbulent
  Clouds} 
In circumgalatic environments stable against Jeans collapse,
turbulence and radiative cooling become the primary drivers
of turbulent cloud formation. Turbulence drives density
fluctuations in the medium, which in turn overcool and
trigger condensation \citep{1996ApJ...469..589M,
  2004MNRAS.355..694M, 2016MNRAS.462.4157A,
  2023ApJ...955L..25C}, forming dense clouds. For strong
turbulence driving we observe this to be hierarchical, with
further condensation and denser clumps formining within
exisiting overdensities. Such clouds remain stable via a
strong pressure gradient between regions internal and
external to the cloud, where the hot high pressure external
region spatially confines the cool low pressure cloud
\citep{2020MNRAS.492.1841L}. Turbulence driving maintains
the strong pressure gradient, where kinetic energy is
continuously, uniformly and isotropically injected into our
box, which cascades into thermal energy and heats the
medum. Dense clouds remain cool at around $10^4$ K since
they can efficiently thermally dissipate the additional
energy, but the diffuse medium must climb to higher
temperatures of $10^5 - 10^6$ K, where $\Lambda$ reaches its
local maxima depending on $Z/Z_\odot$, in order to reach
thermal equilibrium with turbulence driving (the run becomes
subsonic if thermal equilibrium cannot be reached at those
local
maxima). \citet{2018MNRAS.480L.111G,2020MNRAS.492.1970G}
showed that stable clouds can grow via entraining and mixing
with hot gas on its boundary layers. However, those findings
are based in idealized wind tunnel simulations, and we
cannot confirm whether this is present in our simulations
since our "winds" are effectively isotropic.

Our clouds are eventually "crushed" via cooling of the
intracloud medium, which is consistent with
\citet{2020MNRAS.492.1841L}. From Figure
\ref{densmacharray}, the diffuse gas cools significantly
after $10t_s$ for the
$\dot{\epsilon}_i = 10 \; \mathrm{erg} \; \mathrm{s}^{-1} \;
\mathrm{g}^{-1}$, representing the end of the rapid thermal
dissipation epoch as seen in Figure
\ref{totaltimeevolution}. Figures \ref{clumpcompare} and
\ref{powlawvtime} show trends only with respect to $t$ as
opposed to dimensionless time such as $t/t_s$. The $n$
dependence arises from enhanced cooling stabilizing clouds
against dissipation, although it is only present for
$t_\mathrm{Diss}^C$ and absent for $t_\mathrm{cc}$. The
independence of both timescales from $\dot{\epsilon}_i$
raises interesting questions on the interdependencies
between $\chi$, $a_0$ and $v_b$ in eq,~\eqref{eqn:cloudcrush}. From visual inspection,
$\dot{\epsilon}_i$ shares an inverse relation with $a_0$,
but a positive relation with $v_b$ and $\chi$, which may end
up constraining $t_\mathrm{cc}$. A future study focused on
the formation and dissipation of turbulence driven clouds,
particularly in non-idealized environments such as ours, is
highly warranted.

\subsection{Physical Implications of the Density
  Homogenization and Cloud Crushing Timescales}

The density homogenization timescales of $30-300$ Myr and
cloud crushing timescales of 10 Myr independent of
turbulence driving strength may have multiple physical
implications. On the brevity of such timescales, warm and
cool clouds are ubiquitously observed in the CGM
\citep{2013ApJ...763..148S}, while the dynamical time on
circumgalactic scales is several hundred Myr
\citep{1991ApJ...370...78H} and freefall time up to a Gyr,
several times longer than the structural dissipation
timescales. While we neglected microphysics such as
conduction which may play an important role in cloud
stabilization (\citep{1965RvPP....1..205B,
  2020MNRAS.492.1841L}, a more likely explanation would be
periodic feedback-driven outflows from the galactic disc
(\citep{2014ApJ...794..156R, 2015ApJ...812...83N,
  2013ApJ...768...18B} with hot outflow gas accelerating and
shocking the cool CGM gas \citep{2016MNRAS.455.1830T},
though \citet{2015MNRAS.454.2691M} found galactic outflows
to be bursty with periods of up to hundreds of Myr in the
FIRE simulations \citep{2014MNRAS.445..581H}. On the other
hand, feedback from galactic nuclear regions occur on much
shorter timescales of 10s of Myr, particularly for barred
galaxies \citep{2015MNRAS.453..739K}, from which the
feedback of nuclear star clusters are capable of forming
massive biconical outflows \citep{1998MNRAS.293..299T,
  2005ApJ...628L..13T, 2020ApJ...895...43S}. A possible
alternate or compounding explanation is that haloes
exhibiting such clouds are unvirialized and hence see
turbulence driven by supersonic cold streams
\citep{2021ApJ...911...88S}. On the independence of such
timescales relative to $\dot{\epsilon}_i$ may also serve as
a method of ascertaining when the last perturbation occured
regardless of driving strength, from starburst/AGN driven
outflows to accretion flows. However, the positive $n$
dependance of the density homogenization timescales would
require a proper spatial characterization of CGM densities,
which is a challenging task given the complex multiphased
nature of the CGM \citep{2017ARA&A..55..389T}.

\subsection{Future Work}
While the non self-gravitating and optically-thin approximations are valid 
for our diffuse runs, they become less accurate for our denser runs of $n \geq 1 \; \mathrm{cm}^{-3}$ 
where our conditions begin to approach those of the ISM. More detailed exploration
of these denser phases, with consistent thermochemistry and self-gravity, would serve
as a bridge between our study and existing studies on turbulence dissipation in GMCs. Additionally,
while magnetic fields are secondary to radiative cooling in stabilizing clouds against crushing, 
they play an important role in GMCs. We would recommend the inclusion of magnetic fields 
in subsequent studies of these denser phases, or even subsequent studies of the diffuse phases
for the sake of completion.

The highly nonlinear thermal dissipative behaviours highlight the impact of kinematics, compression and overdense clouds
on thermal dissipation. Our energy dissipation timescales and curves can be applied in larger scale cosmological
simulations in lieu of conventional cooling curves, where those overdense clouds we observe in this study cannot
be explicitly resolved. 

\section{Conclusion}
\label{sec:discussion}

 This work explores the saturation and dissipation of
  hydrodynamic turbulence with a large array of 252
  simulations, covering and extending beyond the typical range of physical
  parameters describing typical conditions within the CGM. The
  conclusions are summarized as follows.
\begin{enumerate}
\item Energy dissipation from uniformly driven turbulence
  can be characterized into subsonic and supersonic
  categories depending on the compressibility of the
  gas. Supersonic dissipation sees an initial rapid epoch of
  thermal-dominated dissipation followed by a slower epoch
  of energy dissipation, while subsonic dissipation only
  sees slow dissipation. Thermal dissipation occurs rapidly
  within a few kinetic saturation times before plateauing,
  while kinetic dissipation follows a consistent exponential
  curve.
\item Thermal dissipation occurs via enhanced cooling, with
  a highily nonlinear but strongly positive correlation with
  turbulence driving strength which creates overdense clouds
  via shocks. Kinetic dissipation occurs via the energy
  cascade, and positively but more weakly couples to
  turbulence driving strength which increases the initial
  kinetic energy.
\item Substructure formation is observed in supersonic runs,
  with clumping factors ranging from $2 - 10$ depending on
  turbulent driving strength and the density field spanning
  a few orders of magnitude. Subsonic turbulence sees mostly
  uniform gas, with clumping factor within $10^{-3}$ of 1
  and the density field spanning less than single order of
  magnitude. The density homogenization timescale
  $t_\mathrm{Diss}^C$, defined to be how long it takes for a
  supersonic run to become indistinguishable from a subsonic
  run, falls within the same order of magnitude across all
  runs at around $30-300$ Myr depending on initial density
  but independent of turbulence driving strength.
\item Stronger turbulence driving yields denser, more
  concentrated and more compressible clouds, with flattened
  matter and $\mathcal{M}$ power spectra in the linear
  range. Cloud crushing timescales, defined using the power
  indices of the power spectra, are 10 Myr regardless of
  turbulence driving, and unlike density homogenization
  timescales, are independent from $n$.
\end{enumerate}

\begin{acknowledgments}
This work was supported by the National Science Foundation
of China (11991052, 12233001), the National Key R\&D Program
of China (2022YFF0503401), the China Manned Space Project
(CMS-CSST-2021-A04, CMS-CSST-2021-A06), and the Zhejiang
Laboratory (K2022PE0AB01). Renyue Cen is supported in part by
the National Key Research and Development Program of China.
\end{acknowledgments}

\appendix

\section{Resolution Dependance}
\label{sec:appendix:res}

\subsection{Energy Dissipation}

\begin{figure}[h!]
    \centering
    \incgraph{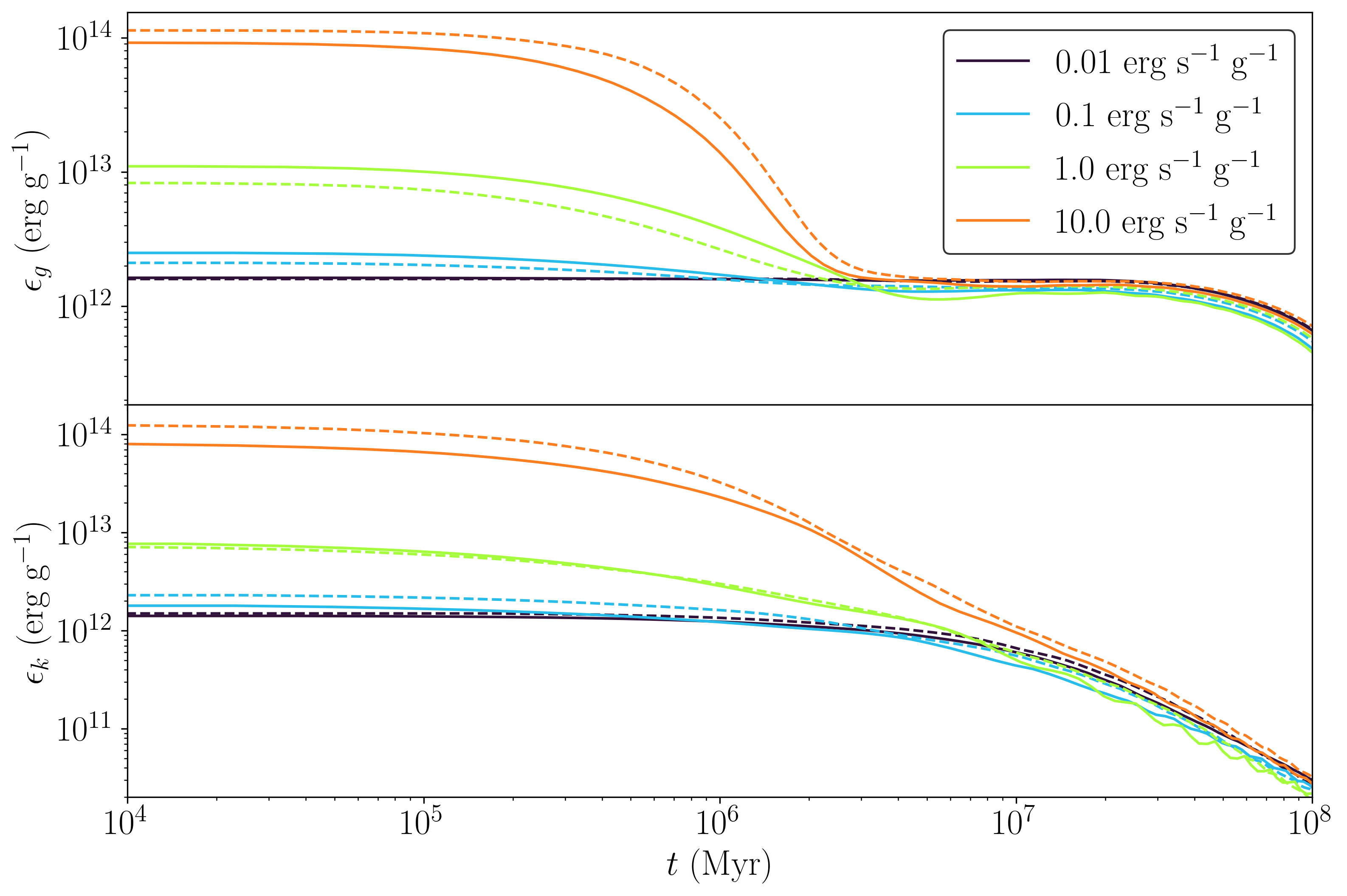}
    \caption{Thermal and kinetic dissipation curves
      comparisons between the fiducial high resolution
      $256^3$ runs (solid line) and lower resolution $128^3$
      counterparts (dashed line). $Z/Z_\odot = 0.3$ and
      $n = 0.1 \; \mathrm{cm}^{-3}$ for these runs.}
    \label{resdep}
\end{figure}

\begin{table} [h!]
  \centering
  \begin{minipage}[b]{\linewidth}
    \centering
    \begin{tabular}{ccc}
      \hline
      $\dot{\epsilon}_i \; (\mathrm{erg} \;
      \mathrm{s}^{-1} \; \mathrm{g}^{-1})$
      & $128^3~t_h$  (Myr)
      & $256^3~t_h$ (Myr)\\ 
      \hline
      \multicolumn{3}{c}{Total Energy} \\
      \hline
      0.01 & 36.0 & 37.4 \\
      0.1 & 4.71 & 3.74 \\
      1 & 0.548 & 0.563 \\
      10 & 0.524 & 0.431 \\
      \hline
      \multicolumn{3}{c}{Thermal Energy} \\
      \hline
      0.01 & 87.5 & 84.8 \\
      0.1 & 51.0 & 23.9 \\
      1 & 0.510 & 0.599 \\
      10 & 0.592 & 0.432 \\
      \hline
      \multicolumn{3}{c}{Kinetic Energy} \\
      \hline
      0.01 & 8.25 & 7.64 \\
      0.1 & 2.49 & 3.24 \\
      1 & 0.623 & 0.502 \\
      10 & 0.434 & 0.425 \\
      \hline
    \end{tabular}
    \caption{Comparisons between half energy dissipation
      timescales $t_h$ between the fiducial high
      resolution $256^3$ runs and their respective lower
      resolution $128^3$ counterparts.}
	\label{restable}
  \end{minipage}
\end{table}

We compare the dissipation curves between our high
resolution runs and their low resolution counterparts in
Figure \ref{resdep}. The lower resolution runs show slightly
different initial energies with variations on order unity
between $0.8$ and $1.5$ times the initial thermal and
kinetic energies of the high resolution runs. The resolution
difference manifests itself primarily as slight differences
in the turbulent steady state, with the dissipation curves
themselves showing near congruent behaviours. Qualitatively
there is good resolution convergence. However, quantitative
differences in the initial states leads to noticeable (and
in some cases significant) quantitative differences in the
dissiption timescales as seen in Table \ref{restable}. The
largest disparity can be seen in the thermal energy
dissipation timescales, where for
$\dot{\epsilon}_i = 0.1 \; (\mathrm{erg} \; \mathrm{s}^{-1}
\; \mathrm{g}^{-1})$ $t_h$ varies by more than a factor of
two. This is the result of the temperature dependence of the
cooling curve, particularly at temperatures around $10^4$ K
where $\dfrac{d\Lambda}{dT}$ is particularly
steep. Disparities in the kinetic dissipation timescales are
less significant, and may result from the finer resolution
scales underestimating the energy cascading rate due to our
lack of subgrid viscosity. Increased resolution leads both
to an increase or decrease in initial energies depending on
driving strength, which suggests that resolution has a
highly nonlinear effect on the initial conditions.

As a comparison, \citet{1998ApJ...508L..99S} saw variations
in total (kinetic + magnetic) energy of around 6\% between
their $128^3$ runs and their $256^3$ runs, which is
significantly less than our variations of up to 50\% in
kinetic energy. This highlights the additional resolution
sensitivity that emerges from an adiabatic equation of
state, where kinetic energy is dissipated via the energy
cascade as opposed to artificial numerical viscosity as
implemented in ZEUS \citep{1992ApJS...80..791S}, which is
the code used by \citet{1998ApJ...508L..99S}. We emphasize
however that there is resolution convergence on the
qualitative conclusions we make on the dissipation curves.

\subsection{Power Spectra}
\begin{figure*}
  \centering
  \incgraph{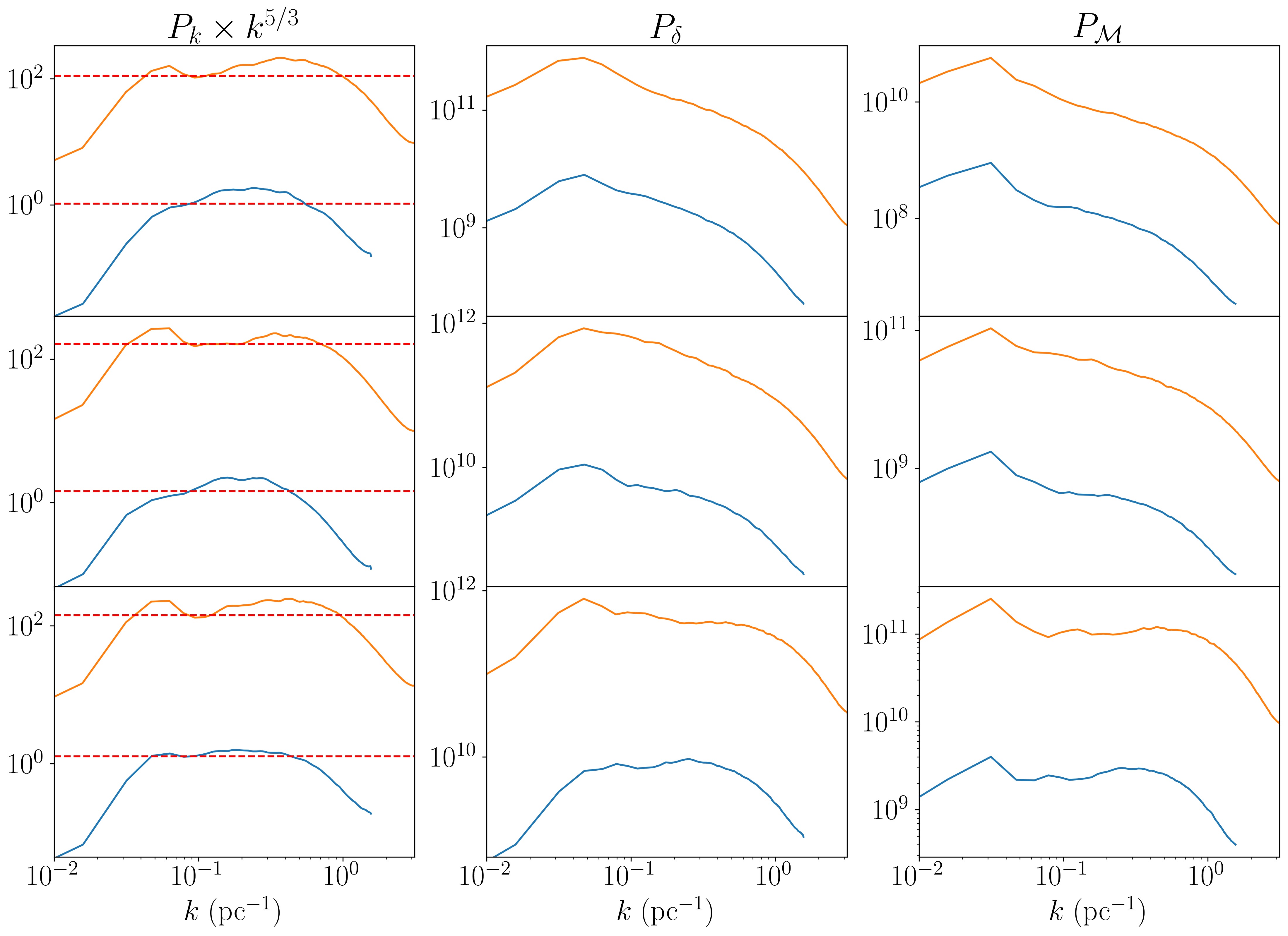}
  \caption{Comparisons between the power spectra of
    $128^3$ (blue) and $256^3$ (orange) resolution
    runs. From left to right we show the turbulence
    ($P_k$), matter ($P_\delta$), and mach
    ($P_\mathcal{M}$) power spectra. $P_k$ is multiplied
    by the $k^{5/3}$ such that kolmogorov power law (red
    dotted) is a horizontal line. From top to bottom we
    have $\dot{\epsilon}_i = 1$, 3 and 10
    $\mathrm{erg} \; \mathrm{s}^{-1} \; \mathrm{g}^{-1}$.}
  \label{psresdep}
\end{figure*}

We present the resolution comparisons of the power spectra
in Figure \ref{psresdep}. The $128^3$ runs are shifted down
by approximately 2 dex and have smaller $k_d$ compared to
the $256^3$ runs. However, we did not utilize or consider
the absolute amplitudes of the power spectra in our
analysis, and $k_d$ was determined dynamically through a
series of curve fits for each run at each snapshot when
fitting $\alpha_\mathcal{M}$ and $\alpha_\delta$. Beyond
these differences, the shapes and trends are broadly
consistent, with both $P_\delta$ and $P_\mathcal{M}$ for
both resolutions seeing a flatter linear range with higher
$\dot{\epsilon}_i$.

The bottleneck effect, which decribes the accumulation of
energy at high $k$ in the inertial regime of the turbulence
power spectrum, sees a dependence on both the
compressibility of the gas ($\mathcal{M}$) as well as the
resolution, being particularly relevant for higher
resolution and strongly supersonic simulations
\citep{PhysRevE.68.026304, PhysRevE.70.016308,
  2010A&A...512A..81F}. We find that the bottleneck effect
remains visible for both $128^3$ and $256^3$ runs, though
strong compressibility does dampen it for the $128^3$
resolution run in the bottom left plot.

\bibliography{Dissipation}{}
\bibliographystyle{aasjournal}

\end{document}